\documentclass[aps,prd,nofootinbib,twocolumn,superscriptaddress,preprintnumbers,balancelastpage,longbibliography]{revtex4-1}

\usepackage{placeins}
\usepackage{amsmath,amssymb,mathtools,bm}
\usepackage{lipsum}
\usepackage{comment}
\usepackage{gensymb}
\usepackage{physics}
\usepackage{cancel}

\usepackage{tablefootnote}
\usepackage{graphicx, color, hepunits}
\usepackage[dvipsnames]{xcolor}
\usepackage{float}
\usepackage{filecontents}
\usepackage{multirow}
\usepackage{slashed}
\usepackage{pifont}

\usepackage[colorlinks=true % false: boxed links; true: colored links
,urlcolor=purple % color of external links
,anchorcolor=blue
,citecolor=blue % color of links to bibliography
,filecolor=magenta  % color of file links
,linkcolor=blue % color of internal links
,menucolor=blue
,linktocpage=true
,pdfproducer=medialab
,pdfa=true
]{hyperref}

\usepackage[utf8]{inputenc}
\usepackage[english]{babel}
\usepackage{soul}
\usepackage{hyperref}

\newcommand{\mpl}{M_{\rm pl}}

\newcommand{\es}[2] {\begin{equation} \label{#1} \begin{split} #2 \end{split} \end{equation}}

\restylefloat{table}

\begin{document}

\title{
String Theory and Grand Unification Suggest a Sub-Microelectronvolt QCD Axion}

\author{Joshua N. Benabou}
\affiliation{Theoretical Physics Group, Lawrence Berkeley National Laboratory, Berkeley, CA 94720, U.S.A.}
\affiliation{Berkeley Center for Theoretical Physics, University of California, Berkeley, CA 94720, U.S.A.}

\author{Katherine Fraser}
\affiliation{Theoretical Physics Group, Lawrence Berkeley National Laboratory, Berkeley, CA 94720, U.S.A.}
\affiliation{Berkeley Center for Theoretical Physics, University of California, Berkeley, CA 94720, U.S.A.}

\author{Mario Reig}
\affiliation{Rudolf Peierls Centre for Theoretical Physics, University of Oxford, Parks Road, Oxford OX1 3PU, UK}

\author{Benjamin R. Safdi}
\affiliation{Theoretical Physics Group, Lawrence Berkeley National Laboratory, Berkeley, CA 94720, U.S.A.}
\affiliation{Berkeley Center for Theoretical Physics, University of California, Berkeley, CA 94720, U.S.A.}

\date{\today}

\begin{abstract} 
Axions, grand unification, and string theory are each compelling extensions of the Standard Model. We show that combining these frameworks imposes strong constraints on the QCD axion mass. Using unitarity arguments and explicit string compactifications—such as those from the Kreuzer–Skarke (KS) type IIB ensemble—we find that the axion mass is favored to lie within the range $10^{-11} \,\text{eV} \lesssim m_a \lesssim  10^{-8} \,\text{eV}$. This range is directly relevant for near-future axion dark matter searches, including ABRACADABRA/DMRadio and CASPEr. 
We argue that grand unification and 
the absence of proton decay suggest a compactification volume that keeps the string scale above the unification scale ($\sim$$10^{16}$ GeV), which in turn limits how heavy the axion can be.  The same requirements limit the KS axiverse to have at most $\sim$47 axions. As an additional application of our methodology, we search for axions in the KS axiverse that could
explain the recent Dark Energy Spectroscopic Instrument (DESI) hints of evolving dark energy but find none with high enough 
decay constant ($f_a \gtrsim 2.5 \times 10^{17}$ GeV); we comment on why such high decay constants and low axion masses are difficult to obtain in string compactifications more broadly. 
\end{abstract}
\maketitle

\section{Introduction}
\label{sec:intro}
The quantum chromodynamics (QCD) axion is a new physics candidate that may explain the Strong {\it CP} problem of the neutron electric dipole moment and also explain the dark matter (DM) of the universe~\cite{Peccei:1977hh,Peccei:1977ur,Weinberg:1977ma,Wilczek:1977pj,Preskill:1982cy,Abbott:1982af,Dine:1982ah}.  The axion is a compact field with a period an integer multiple of $2 \pi f_a$, 
where $f_a$ is the decay constant (see~\cite{Hook:2018dlk,DiLuzio:2020wdo,Safdi:2022xkm,OHare:2024nmr} for reviews).  In the original field theory constructions of the QCD axion it emerges as the pseudo-Goldstone boson of a $U(1)_{\rm PQ}$ symmetry that is spontaneously broken at a scale of order $f_a$; the axion would be exactly massless but for the chiral anomaly, which explicitly breaks the $U(1)_{\rm PQ}$ symmetry and gives rise to a QCD-induced axion mass $m_a \approx 5.7 \times 10^{-6} (10^{12} \, \, {\rm GeV} / f_a)$ eV~\cite{diCortona:2015ldu}.  On the other hand, field theory axion constructions suffer from the Peccei-Quinn (PQ) quality problem~\cite{Georgi:1981pu,Lazarides:1985bj,Kamionkowski:1992mf, Ghigna:1992iv, Barr:1992qq, Holman:1992us,Lu:2023ayc} related to the expectation that quantum gravity should violate global symmetries. String theory, or more generally extra-dimensional constructions, provide the leading dynamical mechanism at present to naturally generate high-quality QCD axions~\cite{Witten:1984dg,Choi:1985je,Barr:1985hk,Svrcek:2006yi,Arvanitaki:2009fg}.  In particular, zero modes of  higher-form gauge fields, which typically arise from the closed string spectrum, compactified on the extra dimensions required by the theory behave like axions and have high quality because of the gauge invariance of the higher-dimensional theory or, equivalently, the generalized global symmetries they possess~\cite{Reece:2024wrn,Craig:2024dnl,Agrawal:2024ejr}.  In these constructions, the axion decay constant is determined by the geometry and volume of the compact manifold.

\begin{figure}[!ht]
    \centering     \includegraphics[width=0.47\textwidth]{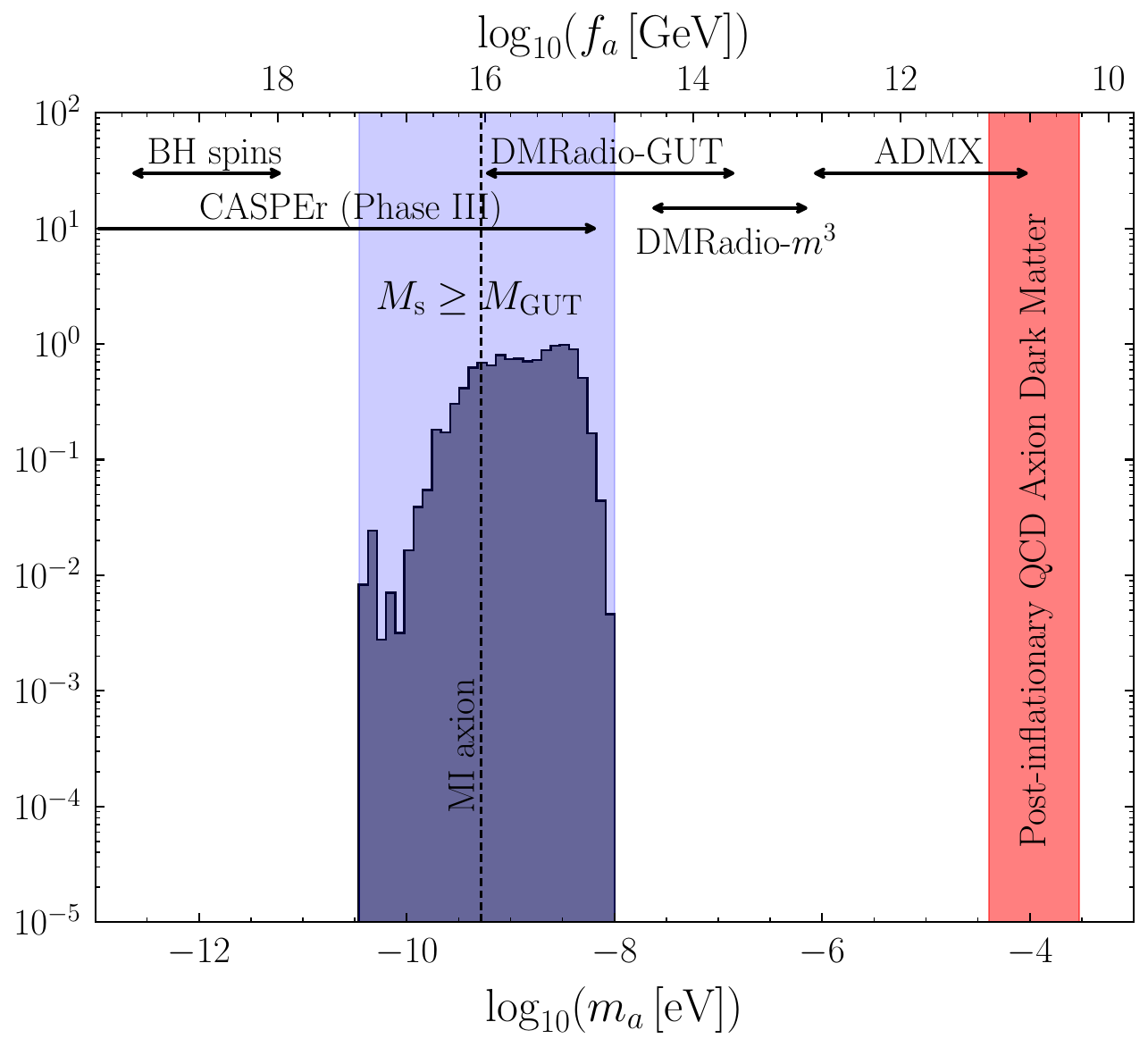}
    \caption{The distribution of the QCD axion mass $m_a$ 
    in the Kreuzer-Skarke (KS) axiverse, restricting to compactifications for which $M_\mathrm{s} \gtrsim 2 \times 10^{16}$ GeV (histogram, see text for details).  The minimal and maximal non-zero values of this histogram are prior independent and bound the possible axion masses consistent with proton decay and grand unification in these constructions.  Note that lower QCD axion masses are disfavored in string theory, as we discuss in the text. 
    Regions targeted by direct detection experiments and searches for black hole superradiance are indicated by arrows, while in red we illustrate the axion mass range where QCD axions could explain the DM abundance by the post-inflationary axion string production mechanism.  As a reference, we also show the mass of the \textit{model-independent} (MI) axion of heterotic string theory.
    }
    \label{fig:money}
\end{figure}

In this work, we show that requiring a grand unified theory (GUT)~\cite{Georgi:1974sy,Fritzsch:1974nn} in the ultraviolet (UV)\textemdash and, in some cases, forbidding proton decay from stringy contributions\textemdash constrains the decay constant in these constructions to lie within a certain range, with dramatic implications for direct-detection experiments (see Fig.~\ref{fig:money}).

The basic intuition behind our result is straightforward. Imagine a 5D construction, where spacetime is described by $M_4 \times S_1$, with $M_4$ the regular 4D Minkowski spacetime and $S_1$ a circle of radius $R$, which may be quotiented by a discrete group in more realistic orbifold constructions~\cite{Kawamura:1999nj,Kawamura:2000ev,Hall:2001pg,Hall:2002ci,Hebecker:2001wq,Altarelli:2001qj}, as we discuss below.  Imagine that the 5D bulk gauge theory is a GUT, which is, for example, broken by the orbifold boundary conditions to the Standard Model (SM). Then, neglecting gauge couplings and factors of order unity that we account for later in this work, we expect the Kaluza-Klein (KK) scale $1/R = M_{\rm KK} \sim M_{\rm GUT}$,  
with $M_{\rm GUT}\sim 10^{16}$ GeV the mass scale associated with grand unification. Since the decay constant is also determined geometrically, $f_a \propto 1/R$, $m_a$ is on the order of a fraction to a few nano-eV.  In particular, in this construction the QCD axion mass is inescapably coupled to the scale of grand unification, since the size of the extra dimension determines both quantities.  Larger axion masses are incompatible with grand unification in this construction because, numerically, the SM gauge couplings do not unify, in supersymmetry (SUSY) or non-SUSY UV completions, at lower mass scales. If lower-scale unification was engineered through {\it e.g.} threshold corrections 
then, in general, one would need to contend with proton decay.  

We show that there is an upper bound on the QCD axion mass in broad classes of string theory constructions, where the axion emerges from closed string modes, that are consistent with grand unification.  Imagine a 10D string theory that has a $p$-form gauge field $C_p$ compactified on a reasonably isotropic 6D compact manifold of total volume ${\mathcal V}_6$.  Integrating $C_p$ over a $p$-cycle 
gives rise to an axion in the 4D effective field theory (EFT) with a decay constant $f_a \sim \mpl / \mathcal{V}_6^{p/6}$, where $M_{\rm pl}$ is the reduced Planck mass and $\mathcal{V}_6$ is the volume of the compact space in string-length units.  
On the other hand, the string scale is set by the total volume as: $M_s \propto M_{\rm pl} / \sqrt{ \mathcal{V}_6}$.  As we increase $\mathcal{V}_6$ to lower the axion decay constant we also necessarily lower the string scale, which eventually becomes smaller than the GUT scale, $M_s<M_{\rm GUT}$.

There are multiple issues with lowering the KK and string scales below $\sim 10^{16}$ GeV. From the phenomenological side, a KK scale below the GUT scale, $M_{\rm KK} \ll M_{\rm GUT}$, alters the standard four-dimensional logarithmic running of gauge couplings, generically lowering the apparent unification scale compared to the MSSM-like prediction~\cite{Dienes:1998vg,Hall:2001pg,Contino:2001si,Agashe:2002pr,Hebecker:2002vm,Hebecker:2004xx,Kumar:2018jxz}.\footnote{Threshold corrections from winding string modes, and their relation to the GUT scale, have also been studied in~\cite{Conlon:2009qa}.
} (In some of the most well-studied examples the KK scales and GUT scales coincide~\cite{Hall:2001pg,Hebecker:2001wq,Kawamura:1999nj,Kawamura:2000ev,Altarelli:2001qj,Witten:1985xc}.) A reduced unification scale not only compromises the precision of SUSY gauge coupling unification near $10^{16}$ GeV, but also tends to lead to enhanced proton decay rates. In minimal SUSY GUTs, predicted decay rates become generically inconsistent with current bounds if $M_{\rm GUT}$ is below $\sim$$10^{16}$ GeV~\cite{Langacker:1980js,Senjanovic:2009kr,Hisano:2022qll,Ohlsson:2023ddi}. 
(Future Hyper-Kamiokande data may strengthen the lower bound on $M_\mathrm{GUT}$ by a factor of  $\sim$2 \cite{Hyper-Kamiokande:2018ofw}.) 
These concerns extend to string theory models in which four-dimensional field-theoretic unification is not realized but unification emerges only in the higher-dimensional theory~\cite{Klebanov:2003my}. Even in intersecting D-brane constructions in type IIA/IIB, where different gauge groups arise on distinct cycles, heavy string modes and stringy instanton effects can generate dangerous dimension-five and dimension-six baryon-number-violating operators, suppressed by the string scale $M_s$.  (As we discuss further below, the string and KK scales are typically separated only by a factor of $\mathcal{O}(1 - 10)$.) Consequently, as we will see later in this work, lowering the string scale well below $10^{16}$ GeV can lead to conflict with proton decay constraints, even in the absence of explicit GUTs, unless specific model-building mechanisms are invoked  
to suppress such operators, {\it e.g.}, involving localization of fields in different places in the compact space~\cite{Arkani-Hamed:1999ylh,Aldazabal:2000cn,Gherghetta:2000qt}. This risk is especially acute given that string theory does not respect global symmetries; baryon and lepton number are expected to be violated by quantum gravitational and stringy effects.
 
We find in a wide range of weakly-coupled type IIB string theory compactifications with ${\mathcal N} = 1$ SUSY, which can be on highly anisotropic manifolds, that 
\es{eq:main}{
3 \times 10^{-11} \, \, {\rm eV} \lesssim m_a \lesssim 10^{-8} \, \, {\rm eV} \,,
}
after imposing the requirement that $M_s$ be above the SUSY GUT scale, which---as we discuss more below---we fix to be $M_{\rm GUT} = 2 \times 10^{16}$ GeV. 
To derive~\eqref{eq:main} we use the Kreuzer-Skarke (KS) \href{http://hep.itp.tuwien.ac.at/%7Ekreuzer/CY/CYcy.html}{database}  of Calabi–Yau (CY) manifolds~\cite{Kreuzer:2000xy}. 
In particular, we compactify the theory on O3/O7 orientifolds of CY 3-fold hypersurfaces of toric varieties. The CY 3-folds are obtained by triangulating reflexive polytopes of dimension 4 following the construction of Batryev~\cite{Batyrev:1993oya}. All 473,800,776 such polytopes, which extend to $h^{1,1}=491$, have been enumerated in the KS database~\cite{Kreuzer:2000xy}. The efficient computation of triangulations for Hodge numbers $h^{1,1} \gg 1$ has only recently been made possible by algorithms introduced in Refs.~\cite{Demirtas:2018akl,Demirtas:2022hqf}, with a publicly available implementation introduced in~\cite{Demirtas:2022hqf} in the form of the \href{https://cy.tools/}{\texttt{CYTools}} package, which we use in this work.

In these theories, the Hodge number $h^{1,1}$ gives the number of axion-like particles arising from the Ramond-Ramond (RR) four-form $C_4$; these are the closed-string axions that can couple to SM gauge groups.  It has been previously observed that increasing $h^{1,1}$ correlates with decreasing $f_a$~\cite{Gendler:2023kjt} (also in $F$-theory constructions~\cite{Halverson:2019cmy}).  As such, it is not surprising that we find that there is an upper limit on $h^{1,1}$ for manifolds that are compatible with SUSY GUTs and, more generally, proton decay;  
in particular, we find that only manifolds with $h^{1,1} \lesssim 47$ are allowed.  
This result has implications for phenomenology  in the axiverse~\cite{Arvanitaki:2009fg,Svrcek:2006yi,Cicoli:2012sz}. 

While we derive~\eqref{eq:main} in the specific context of weakly coupled type IIB string theory, we expect it to also hold true in other broad regions of the string landscape.  We show this through partial-wave unitarity arguments, which imply that $m_a \lesssim 6 \times 10^{-8}$ eV is required in order for the 4D QCD axion EFT to be unitary at energy scales up to $M_{\rm GUT}$.   

There are a few ways around the axion mass range prediction in~\eqref{eq:main} within the context of string theory axions.  For example, as recently considered in~\cite{Choi:2011xt,Cicoli:2013ana,Cicoli:2013cha,Buchbinder:2014qca,Choi:2014uaa,Allahverdi:2014ppa,Petrossian-Byrne:2025mto,Loladze:2025uvf}, it is possible to separate the relation between the string scale and the axion decay constant through a non-geometric axion implementation, as happens for example in cases where the axion arises from pseudo-anomalous $U(1)$ symmetries in superstring constructions~\cite{Dine:1987xk}, analogously to field theory axion constructions (though obtaining parametrically low axion decay constants below $M_{s}$ may involve tuning in moduli space).  Secondly, even for closed-string-sector axions it may be possible to achieve $f_a \ll M_s$ through warped compactifications, as we discuss later in this work. However, in this case $M_{\rm KK}$ is warped similarly to $f_a$, so one must give up on precision SUSY gauge unification 
and implement a localization mechanism  
to suppress higher-dimensional proton-decay-generating operators~\cite{Gherghetta:2000qt}.  We will also discuss warping in the context of heterotic M-theory, where $M_{11}$ decreases as we lower $f_a$.

The mass range prediction in~\eqref{eq:main} is highly relevant to experimental probes of axion DM, as it is outside the $\sim$1–1000 $\mu$eV mass range targeted by a number of terrestrial experiments such as ADMX~\cite{Du:2018uak,Braine:2019fqb,ADMX:2018gho,ADMX:2019uok,ADMX:2020hay}, HAYSTAC~\cite{Zhong:2018rsr,Backes:2020ajv,HAYSTAC:2020kwv}, MADMAX~\cite{Caldwell:2016dcw,MADMAXinterestGroup:2017koy,MADMAX:2019pub,Garcia:2024xzc}, ALPHA~\cite{Lawson:2019brd,Wooten:2022vpj,ALPHA:2022rxj}, BREAD~\cite{BREAD:2021tpx,BREAD:2023xhc}, and DALI~\cite{DeMiguel:2023nmz,DeMiguel:2020rpn}, among others~\cite{Adams:2022pbo}. 
(Note, however, that searching in the ${\mathcal O}(10 - 100)$ $\mu$eV subset of this mass range is motivated by the post-inflationary axion cosmology, with cosmological axion strings and domain walls producing the DM relic abundance \cite{Klaer:2017ond,Gorghetto:2018myk,Vaquero:2018tib,Drew:2019mzc,Drew:2022iqz,Drew:2023ptp,Gorghetto:2020qws,Dine:2020pds,Buschmann:2021sdq,Kim:2024wku,Saikawa:2024bta,Kim:2024dtq,Benabou:2024msj}.)
Similarly, our preferred region is outside that currently probed by black hole superradiance~\cite{Arvanitaki:2009fg,Arvanitaki:2010sy} (see  \cite{Baryakhtar:2020gao,Witte:2024drg} for recent studies including self-interactions), which presently excludes QCD axions with masses below those we find in explicit string constructions. 
On the other hand, our work strongly motivates the pursuit of axion DM experiments in the mass range given in~\eqref{eq:main}, such as the lumped-element ABRACADABRA experiment~\cite{Kahn:2016aff,Ouellet:2018beu,Ouellet:2019tlz,Salemi:2021gck} (and the follow-up DMRadio program~\cite{DMRadio:2022pkf,DMRadio:2022jfv,Benabou:2022qpv,DMRadio:2023igr}), related proposals using superconducting resonant frequency cavity conversion~\cite{Berlin:2019ahk,Giaccone:2022pke}, and axion-nucleon spin-precession experiments such as CASPEr~\cite{Graham:2013gfa,Budker:2013hfa,JacksonKimball:2017elr,Aybas:2021cdk,Dror:2022xpi} (see Fig.~\ref{fig:money}).

\section{Bounding $f_a$ with Partial-wave unitarity}
\label{sec:unitarity}
Before considering explicit examples, we provide evidence for the existence of model-independent lower bounds on $f_a$ for higher-dimensional axions based on partial-wave unitarity. (See~\cite{Reece:2024wrn} for a related approach using renormalizability.) 
Partial-wave unitarity is a powerful way to constrain the strength of effective interactions. The strategy is based on expanding the $2\rightarrow 2$ amplitude in partial-waves and then imposing unitarity conditions on the amplitudes (see, {\it e.g.},~\cite{Jacob:1959at,Lee:1977yc,Lee:1977eg}). This method can be used to obtain bounds on couplings or to estimate the scale at which an EFT breaks down. A well-known example is the case of Chiral Perturbation Theory ($\chi $PT)~\cite{Weinberg:1978kz,Manohar:1983md}, where by analyzing $\pi-\pi$ scattering one can predict the cut-off scale of the theory as a function of the pion decay constant, $f_\pi$. Pions are derivatively coupled pseudo-Goldstone bosons, and the $2\rightarrow 2$ amplitude grows with the Mandelstam variable $s$ as $\mathcal{A}(s)\sim \frac{s}{f_\pi^2}$. From this we can estimate the cut-off scale, $\Lambda_{\chi }\sim 4\pi f_\pi\sim 1.2$~GeV. At this scale $\chi $PT breaks down and new states, including the $\rho$ meson and other resonances, appear. Similar arguments have been used in the case of massive $U(1)$ gauge bosons in chiral theories to link the mass of the Higgs to the mass of the heaviest fermion with $U(1)$ charge~\cite{Craig:2019zkf}. As we show here, this method can also be used to obtain a lower bound on the axion decay constant for higher-dimensional axions. 

If the SM undergoes grand unification then the QCD axion must couple to electroweak gauge bosons in addition to gluons. However, the unitarity constraints arising from gluon scattering are the most constraining because of the large multiplicity of gluons. In particular, the most constraining partial-wave unitarity bound comes from the total angular momentum ${\mathcal J} = 0$ elastic scattering $G_{\pm} G_{\pm} \to G_{\pm} G_{\pm}$ of two gluons $G$ in the color-singlet state, with subscripts denoting helicities, with the scattering mediated by the axion. 
We write the axion-QCD interaction as
\es{eq:axion_QCD}{
{\mathcal L} \supset {\alpha_s \over 8 \pi} {a \over f_a}  G^a_{\mu \nu} \tilde G^{a \, \mu \nu} \,,
}
with $G$ the QCD field strength, $\tilde G$ its dual, and $\alpha_s$ the strong fine-structure constant.  
Given that~\eqref{eq:axion_QCD} is a dimension-5 operator, the gluon elastic scattering rate will clearly violate the partial-wave unitarity bound at a sufficiently high center of mass energy $\sqrt{s}$, since the amplitude for the scattering process must scale as ${\mathcal A}(s) \sim {s \over f_a^2}$, in analogy with that in pion scattering.  
 
Implementing the partial-wave unitarity bound for the gluon-gluon scattering process for the color-singlet state above imposes~\cite{Brivio:2021fog}\footnote{Note that this equation formally requires us to be in the weakly-coupled 't Hooft limit.
} 
\es{eq:partial_wave}{
{4 (N_c^2 - 1) \over \pi} {s  \over f_a^2} {\alpha_s^2 \over 64 \pi^2} \lesssim 1 \,,
}
with $N_c = 3$ for QCD. 
The axion EFT should be valid until the energy scale of new resonances associated with the UV completion.  Equation~\eqref{eq:partial_wave} is not valid at energy scales above the Kaluza Klein (KK) scale, because new resonances appear at that scale for the axion and gluon that enter into the partial-wave unitarity calculation. On the other hand, the theory above the KK scale is still non-renormalizable.  That is, the KK modes that enter at the KK scale are not able to unitarize the theory. In a string theory UV completion compactified on a flat spacetime, string states at the string scale, which is above the KK scale, unitarize the scattering. (In a warped compactification, the theory may become strongly coupled above the KK scale.) We return to these points later, but we require that the scattering amplitude in~\eqref{eq:partial_wave} obeys the unitarity constraint (and possibly saturates the constraint) at the KK scale, which we define---in the case of multiple extra dimensions---as the mass of the lightest massive resonance which propagates through the SM gauge cycle ({\it e.g.}, this is the mass of the lightest KK-excitation of the SM gluons). 

Substituting $s = M_{\rm KK}^2$ into~\eqref{eq:partial_wave} we find that
\es{eq:unitarity_0}{
f_a \gtrsim M_{\rm KK} {\alpha_s(M_{\rm KK} ) \over \sqrt{2 \pi^3}} \,.
}
To find the lowest allowed value of $f_a$ we should consider KK scales all the way down to $M_{\rm GUT}=2\times 10^{16}$ GeV, for the reasons explained in Sec.~\ref{sec:intro}. 
We also fix the GUT scale gauge coupling to the value for SUSY UV completions, 
 $\alpha_s(M_{\rm GUT}) = \alpha_{\rm GUT} = 25^{-1}$, for definiteness.  
Substituting these values into~\eqref{eq:unitarity_0} determines the unitarity lower-bound on the higher-dimensional axion decay constant
\es{eq:bound}{
{f_a} \gtrsim 10^{14} \, \, {\rm GeV} \to m_a  \lesssim 6 \times 10^{-8} \, \, {\rm eV} \,.
}
 
In a four-dimensional field theory UV completion, it is too strong of a requirement to impose that the axion EFT be valid up to $M_{\rm GUT}$. As we show in the next section, in a field theory context it is perfectly acceptable for the axion to appear as a pseudo-Goldstone boson of a symmetry broken at energy scales below $M_{\rm GUT}$. In that case, the radial mode and, possibly, other fermions or scalars with PQ charges, have a mass below $M_{\rm GUT}$ and are responsible for unitarizing the scattering amplitude.
Taking $f_a \ll M_{\rm GUT}$ for a field theory axion does not affect grand unification, apart from possible small modifications to the running of the SM gauge couplings from extra vector-like fermions, and does not affect proton stability.

\section{QCD axion in higher-dimensional Unified Theories}
\label{sec:toy_examples}

In this section we provide examples where a QCD axion mass range can be obtained analytically in the context of closed-string axions or extra-dimensional field theory analogues. We show that in certain situations---including orbifold GUTs, $p-$forms compactified on factorizable manifolds, and heterotic strings---the axion decay constant is fixed by the definition of the gauge coupling. On the other hand, in scenarios where there are extra dimensions through which only closed-string modes such as gravitons propagate, or in scenarios with non-factorizable geometries, the string scale $M_s$ can be decreased by changing the volume of the internal space. In this case the decay constant can be decreased by lowering the fundamental scale, $M_s$, though only by a certain amount.  

\subsection{4D field theory KSVZ GUT model}
\label{sec:field_theory}

Before discussing axions in the context of extra-dimensional GUTs, we briefly review why field theory axions in 4D GUTs may achieve $f_a \ll M_{\rm GUT}$.  For simplicity, we illustrate this point in the context of the  Kim–Shifman–Vainshtein–Zakharov (KSVZ) type axion model~\cite{Kim:1979if,Shifman:1979if}.  As an example, we take the case of an $SU(5)$ non-supersymmetric GUT.  We supplement the standard embedding of the SM into $SU(5)$ with a vector-like Dirac fermion, denoted by $5_F$. (Note that the inclusion of $5_F$ slightly modifies $\alpha_{\rm GUT}$ but not $M_{\rm GUT}$ or the precision of unification.)  We also add to the theory a complex scalar $\Phi$ that is a singlet under $SU(5)$.  Then, consider the Lagrangian 
\es{}{
{\mathcal L} \supset {\mathcal L}_{SU(5)} + {\mathcal L}_{\rm PQ} + \bar 5_F i \slashed{D} 5_F +  ( y \Phi \bar 5_F 5_F + {\rm h.c.}) \,,  
}
where
\es{}{
{\mathcal L}_{\rm PQ} = |\partial \Phi|^2 - \lambda \left( |\Phi|^2 - {v_a^2 \over 2} \right)^2 \,,
}
and ${\mathcal L}_{SU(5)}$ refers to the Lagrangian for the standard embedding of the SM into $SU(5)$, including the scalar sector responsible for breaking $SU(5) \to SU(3) \times SU(2) \times U(1)$.
The PQ symmetry, which is anomalous, acts in the usual way ({\it e.g.}, $\Phi \to e^{i \alpha}\Phi$ and $5_F \to e^{i \alpha \gamma_5 / 2} 5_F$).  The KSVZ axion, which is identified by the phase of $\Phi$, has a decay constant $f_a = v_a$ in this case. (With, instead, $N$ copies of $5_F$ the decay constant would be $f_a = v_a / N$.)
 
The key point is that in standard 4D GUTs we may have $v_a \ll M_{\rm GUT}$ and thus $f_a \ll M_{\rm GUT}$ (see for example~\cite{Ernst:2018bib}), though this may involve fine tuning in some cases. Most naturally, we would expect $v_a \sim M_{\rm GUT}$~\cite{Wise:1981ry,DiLuzio:2018gqe,FileviezPerez:2019fku}, but the presence of, for example, the doublet-triplet splitting problem and the electroweak hierarchy problem tells us that some amount of fine tuning is likely at play in this model anyway.  Note that this tuning is similar to that needed in the case of `higher axions'  
to achieve decay constants much less than $M_{\rm GUT}$~\cite{Choi:2011xt,Cicoli:2013ana,Cicoli:2013cha,Buchbinder:2014qca,Choi:2014uaa,Allahverdi:2014ppa,Petrossian-Byrne:2025mto,Loladze:2025uvf}.

The above example shows that in field theory GUTs the decay constant can, with tuning, be brought below the GUT scale. While there are some field theory GUT axion models for which it is harder to tune the decay constant to be below the GUT scale ({\it e.g.}, those that embed the axion as a phase in the field responsible for GUT symmetry breaking~\cite{Wise:1981ry}, for which the decay constant tends to be linked more closely with the GUT scale), this is the exception for field theory constructions. In contrast, in the remainder of this work, we will show that for string theory UV completions with closed string axions, it is harder to tune the decay constant to be below $M_{\rm GUT}$.

\subsection{QCD axion in a 5D orbifold GUT} 
\label{sec:5D_orbifold_flat}

An orbifold GUT is a higher-dimensional field theory where the extra dimensions are compactified on an orbifold \cite{Kawamura:1999nj,Hall:2001pg,Altarelli:2001qj,Hebecker:2001wq}. These constructions admit defects---that is, branes---at the orbifold singularities. These theories allow for the breaking of the bulk gauge group by orbifold boundary conditions without the need for Higgs fields, solving long-standing problems of 4D GUT theories such as the doublet-triplet splitting problem as well as avoiding wrong relations between fermion masses. The defect at the orbifold singularity admits quantum fields on it. The crucial point is that these fields and their interactions only need to respect the symmetries allowed by orbifold parity and not the bulk gauge symmetry. 

Axions can be incorporated in the framework of orbifold GUTs in close analogy with other 5D constructions \cite{Choi:2003wr} (see \cite{Reece:2024wrn} for a recent review). In addition to the bulk GUT gauge symmetry, one adds an auxiliary $U(1)_A$ gauge symmetry with 5D gauge field $A_M$. We assume that under the orbifold boundary conditions, the 4D part $A_\mu$ is odd while $A_5$ is even, allowing for a zero mode for the latter. In this case the 5$^{\rm th}$ component behaves as the axion: $\theta = \int A_5$. 
This field inherits couplings to 4D gauge bosons from the 5D Chern-Simons (CS) coupling
\begin{equation}
\label{eq:CS}
   \frac{\kappa}{8\pi^2} \int A\wedge \text{tr}\left [G_{\rm GUT}\wedge G_{\rm GUT}\right ]  \rightarrow \kappa \frac{\theta}{{32\pi^2}} \left (\sum_i G_{i}^a\tilde{G}_{i}^a \right ) \,,
\end{equation}
with $\kappa$ the integer-quantized CS level and $i$ indicating the unbroken gauge groups in 4D. 
After going to the canonically normalized form, this yields the expression in~\eqref{eq:axion_QCD}. Since the CS coupling of $A_M$ to the GUT gauge fields is a bulk interaction, in the 4D EFT the axion couples to gauge bosons in a universal way---that is, GUT symmetrically---even if  
the boundary theory does not respect GUT relations. This guarantees that $\theta$ behaves as the QCD axion in close analogy with both the QCD axion in 4D field theory GUTs and heterotic string theory axions with the SM embedded in an $E_8$ factor~\cite{Agrawal:2022lsp,Agrawal:2024ejr}.
In the absence of light matter charged under $U(1)_A$, $A_5$ has exponentially good quality due to the global 1-form symmetry of $A_M$. Any effect breaking the shift symmetry of $\theta$ apart from IR instantons of the GUT symmetry is exponentially suppressed by the action $S$, {\it i.e.} as $V(\theta)\propto e^{-S}$. In a field theoretic approach this action is $S=2\pi RM$, with $M$ the mass of particles charged under the $U(1)_A$.  (We discuss later how in string theory it is natural to expect $R M \sim 1/ \alpha_{\rm GUT}$.)

The axion decay constant is obtained from the kinetic term of the $U(1)_A$ 5D gauge field
\begin{equation}
  S\supset  -\int \frac{1}{2
g_5^2}F_{MN}^2 \rightarrow \int \frac{f_a^2}{2} d\theta^2 \,,
\end{equation}
with $f_a^2 = ( \pi R g_5^2)^{-1}$ and
where $g_5$ is the $U(1)_A$ 5D gauge coupling.\footnote{Technically the $f_a$ appearing in the kinetic term for the periodic axion field is not the same as that in {\it e.g.}~\eqref{eq:axion_QCD}, which controls the axion-QCD coupling; they differ by the domain wall number ($\kappa$ in~\eqref{eq:CS}).  Note that in this section we set $\kappa = 1$ for simplicity. In our explicit KS type IIB constructions the CS level is also one by construction, as a result of assuming $f_a$ defined as in~\eqref{eq:axion_QCD}. More general domain wall numbers, obtained via a higher-level embedding, are described in Appendix \ref{app:Ndw}.} Even though the 4D massless vector field is projected out of the spectrum by the orbifold boundary conditions, we may define a 4D gauge coupling for the KK modes and the axion:
\begin{equation}\label{eq:gauge_coupling_orbifoldGUT}
    \frac{1}{\alpha_{a}}=4\pi\frac{\pi R}{g_{5}^2} \,. 
\end{equation}
This allows us to write the decay constant as 
\es{eq:f_a_Mkk}{
f_a = M_{{\rm KK}} \sqrt{ 1 \over \alpha_a 4 \pi^3} \,,
}
with $M_{\rm KK} = 1/R$ the KK scale. 
Then, identifying $M_{\rm KK}$ with $M_{\rm GUT}$ and approximating $\alpha_a \approx \alpha_{\rm GUT}$,\footnote{As we discuss more below, since both gauge couplings are determined geometrically it is natural to expect $\alpha_a \sim \alpha_{\rm GUT}$.}  
we find $f_a \approx 9 \times 10^{15}$ GeV giving $m_a \sim 7 \times 10^{-10}$ eV. 

Equation~\eqref{eq:f_a_Mkk} shows how $f_a$ is directly given by the volume of the extra dimension, up to dependence on $\alpha_a$.  To gain further insight into the expectation for $\alpha_a$, we may consider the embedding of the orbifold theory into a string theory UV completion. 
We let the SM gauge groups arise from fields localized on D-branes, such that, for example, the part of the 5D effective action for gravity and the GUT gauge field is 
 \es{eq:5D_s}{
S_{5D} = 2 \pi \int d^5x {M_s^3 \over g_s^2} R_{5D} - \int d^5x {M_s \over g_s} {1 \over 16 \pi} \left( G_{\rm GUT}^a \right)^2 \,,
}
with $R_{5D}$ the 5D Ricci scalar and where we assume the D-branes fill all of spacetime. 
Above, we introduce the string coupling $g_s$ and the string scale $M_s \equiv 1 / \ell_s$, with $\ell_s$ the string length.  Implicitly, we assume that all additional dimensions of string theory have already been compactified on compact dimensions of length $\ell_s$, such that we may concentrate on the 5D theory; we generalize beyond this assumption starting in Sec.~\ref{sec:p-form_axion}.

The axion emerges as the zero mode of a bulk (1-form) gauge field $C_1$---if we identify this field as a closed-string ({\it e.g.}, RR) mode then its contribution to the 5D action is
\es{eq:SRR_5D}{
S_\mathrm{RR} = -2 \pi \int M_s {1 \over 2} F_2 \wedge \star F_2 + 2 \pi \int_{S^1} C_1  \,,
}
where $\star$ indicates the Hodge dual, $F_{2}$ is the gauge-invariant field strength of the RR 1-form field, which has CS interactions with the GUT theory as in~\eqref{eq:CS}, and where the last term involves the integral of $C_1$ over the 5$^{\rm th}$ dimension (a circle with radius given by $R$).\footnote{The second term in~\eqref{eq:SRR_5D} can be interpreted as the contribution to the action from the D$0$-brane, with charge $+1$ under $C_1$, whose worldline wraps the $S^1$.}  

Note that we may identify the $2\pi$ periodic axion field by $\theta = 2 \pi \int_{S^1} C_1$, which differs by a $2 \pi$ normalization factor from the definition of the axion in the previous example; this normalization factor is convenient in string theory~\cite{Polchinski:1998rr,Svrcek:2006yi}. This leads to the expression for the axion decay constant 
\es{eq:decay_const_5D_flat_stringy_orbifold}{
f_a^2 = {M_s \over 2 \pi^2 R} \,.
}  
Another important scale we may extract from~\eqref{eq:5D_s} is the 4D reduced Planck mass, which is given by
\begin{equation}\label{eq:PlanckMass_flat_orbifold}
    \mpl^2=\frac{4\pi}{g_s^2}M_s^3 \pi R
    \,.
\end{equation}
Moreover, in this scenario, the GUT gauge coupling is determined by 
\es{}{\label{eq:gauge_coupling_orbifoldGUT}
\frac{1}{\alpha_{\rm GUT}} = \frac{1}{g_s}\frac{\pi R}{l_s} \,.
}
Altogether, we can use the definitions of the 4D GUT gauge coupling and 4D Planck scale to write the decay constant as:
\begin{equation}
    f_a = {\alpha_{\rm GUT} \over \sqrt{8 \pi^2 }} M_{\rm pl}\,.
    \label{eq:f_a_orbifolgGUT}
\end{equation}
Substituting in $\alpha_{\rm GUT} \sim 1 / 25$ we find $f_a \approx 1.1 \times 10^{16}$~GeV and $m_a \approx 5.2\times 10^{-10}$~eV.  This axion decay constant is that of the  `model-independent axion' in string theory (see, {\it e.g.},~\cite{Svrcek:2006yi}).  
Comparing with ~\eqref{eq:f_a_Mkk}, we see that the theory where the axion comes from an RR 1-form field has $\alpha_a  = \alpha_{\rm GUT} / (2 g_s)$ and $M_{\rm KK} = M_s ( \pi \alpha_{\rm GUT} / g_s)$. Note that perturbativity requires $g_s \lesssim 1$, while requiring $\pi R \gtrsim \ell_s$ places a lower bound on the string coupling $g_s \gtrsim  \alpha_{\rm GUT}$.  

The relation in~\eqref{eq:f_a_orbifolgGUT} may be modified in the case of a warped compactification, where the 5D metric is given by~\cite{Randall:1999ee}
\begin{align}
    d s^2=G_{M N} d x^M d x^N= e^{-2 k y} \eta_{\mu \nu} d x^\mu d x^\nu+d y^2 \,,
\end{align}
with $k$ the AdS curvature scale and $y$ taking values in the interval $[0,\pi R]$.
We consider this case in detail in App.~\ref{sec:warped}.  As we show in the Appendix, in the warped case $f_a \propto \sqrt{k M_s} e^{- k \pi R}$. Consider the case $k R \sim \mathcal{O}(1)$ and $R^{-1} < M_s$, as required by \eqref{eq:gauge_coupling_orbifoldGUT}, with the string scale $M_s \gtrsim 10^{16} \, \, {\rm GeV}$.  In this case we achieve $f_a \ll 10^{16} \, \, {\rm GeV}$. (For example, taking $M_s \sim 10^{16} \, \, {\rm GeV}$ we may achieve the phenomenologically interesting decay constant $f_a \sim 10^{12}$ GeV for $k R \sim 3$.)  As we discuss further in App.~\ref{sec:warped}, stringy-induced higher-dimension operators that generate proton decay may be suppressed by $M_s$ in this case through localization, providing an example of how localization can be used to evade our conclusions by achieving a low value of fa without violating proton decay constraints.  
While this is true, we also note that in this case the KK scale is also warped to be of the same order as $f_a$: $M_{\rm KK} \sim k e^{- k \pi R} \sim f_a$. At energy scales above the KK scale the gauge couplings run linearly and not logarithmically, which complicates the predictability of precision gauge unification. (That is, in our $f_a \sim 10^{12}$ GeV example large threshold corrections would be required to achieve a GUT scale near $10^{12}$ GeV, or grand unification must be abandoned.)

Similarly,  
it has been shown that in warped compactifications of heterotic M-theory one can decrease the axion decay constant and bring it closer to the
window $10^9\lesssim f_a\lesssim 10^{12}$ GeV when the SM is embedded into the $E_8$ gauge sector supported on the smaller boundary~\cite{Svrcek:2006yi,Im:2019cnl}. This occurs as a result of lowering the fundamental 11D Planck scale, $M_{11}$, with respect to the 4D Planck scale $\mpl$ through warping. The axion decay constant can be lowered with respect to $\mpl$ and turns out to be proportional to the fundamental scale $M_{11}$:
\begin{equation}
    f_a\sim M_{11}\sqrt{\frac{3\alpha_{\rm GUT}}{4\pi}}  \,.
\end{equation}
Thus, one may decrease the decay constant by lowering $M_{11}$.
On the other hand, it is unclear how the theory behaves at energy scales above $M_{11}$, where 11d supergravity and string theory do not provide an accurate description, and what this implies for proton decay rates and gauge unification. New degrees of freedom become dynamical above the scale $M_{11}$ that could propagate in the 11$^{\rm th}$ dimension, such that the theory will likely look very different from a theory with standard GUT phenomenology.\footnote{Additionally, it is likely that KK modes of the gauge bosons on the 10d boundary become relevant at an energy scale below $M_{11}$, which could cause issue for proton decay.}
While being an interesting avenue to achieve low axion decay constants, we do not consider warped compactifications further in this work because they do not provide standard GUT-like phenomenology when $f_a$ is warped substantially below $\sim$$10^{16}$ GeV.   

\subsection{Axion from $p-$form field}
\label{sec:p-form_axion}

In close analogy with the (unwarped) discussion in the previous section, let us now consider axions coming from $p-$forms, $C_p$, integrated over $p-$cycles in the context of string theory allowing all compact dimensions to be extended beyond $\ell_s$. Gauge sectors come in this case  from $D(p+3)-$branes wrapping $p-$cycles. This case is slightly
different from the orbifold case because the axion comes from a closed string mode which can propagate in 10 dimensions and is not restricted to the visible sector cycle. 
We closely follow the discussion and notation in~\cite{Svrcek:2006yi}.

The axion decay constant can be obtained from the kinetic term of the $p-$form field, which follows from the 10D supergravity action, from which~\eqref{eq:SRR_5D} was also obtained:
\begin{equation}
    S_\mathrm{kin}=-\frac{1}{2}\frac{2\pi}{l_s^{8-2p}}\int_{X_6} F_{p+1}\wedge\star F_{p+1}\,,
\end{equation}
where $F_{p+1}=dC_p$ is the gauge invariant field strength. 
In general, the compact space, which we assume is 6D and which we denote by $X_6$, has $b_p$ (with $b_p$ the Betti number) independent $p$-cycles $[D_\alpha]$  
indexed by $\alpha$, which form a basis of the homology group $H_p(X,\mathbb{Z})$. We can write the decomposition of $C_p$ in a basis of harmonic forms $\{\omega_\alpha\}$ dual to the basis of the homology group, $C_p=\frac{1}{2\pi}\sum a_\alpha \omega_\alpha$.  Each of the $a_\alpha$ acts like a $2\pi$ periodic axion under dimensional reduction to 4D.  However, in this section we restrict to the single axion case, and so we write $C_p = {1 \over 2 \pi} a \omega$.  Note that the harmonic form $ \omega$ is normalized such that $\int_{X_p}  \omega = 1$, with $X_p$ the $p$-cycle giving rise to our axion of interest. We define the volume of $X_p$ to be $M_s^{-p} {\mathcal V}_p$ and the total volume of the compact space, $X_6$, to be $M_s^{-6} {\mathcal V}_6$.  That is, the volumes are dimensionless and represented in string length units. Then, without loss of generality we may write
\es{eq:DwedgeD}{
\int_{X_6} \omega \wedge \star \omega = x M_s^{2p - 6} { {\mathcal V}_{6} \over {\mathcal V}_p^2} \,,
}
where $x$ is a dimensionless coefficient that is a function of the geometry which characterizes the non-factorizability of the manifold. In particular, if the 6D space factorizes as $X_6 = X_p \times X_{6-p}$ then $x = 1$, though in general $x \neq 1$. 

The 4D Planck scale and the 4D GUT gauge coupling are determined through dimensional reduction as
\es{eq:key_relations}{
M_{\rm pl}^2 = \frac{4\pi}{g_s^2} \mathcal{V}_{6} M_s^2 \,,\,\,\,\,\frac{1}{\alpha_{\rm GUT}} = \frac{1}{g_s}\mathcal{V}_{p}\,.
}
From these relations we may then write the axion decay constant as
\es{eq:p_form}{
    f_a=\sqrt{x} \frac{\alpha_{\rm GUT}}{\sqrt{8\pi^2}}\mpl\,.
}
For factorizable manifolds we thus obtain the same axion decay constant prediction as in the 5D example in~\eqref{eq:f_a_orbifolgGUT} and the model-independent heterotic axion.
While in the factorizable case the axion decay constant is independent of the total and gauge cycle volumes, that does not mean that all volumes are allowed. First, let us focus on ${\mathcal V}_p$.  We avoid compactifying on dimensions smaller than the string length (requiring ${\mathcal V}_p > 1$), which sets the restriction $g_s \gtrsim \alpha_{\rm GUT}$. On the other hand, perturbativity requires $g_s \lesssim 1$, leading to a maximum gauge cycle volume ${\mathcal V}_p \lesssim \alpha_{\rm GUT}^{-1}$.  Requiring the string scale to be larger than the GUT scale sets the constraint ${\mathcal V}_6 \lesssim 1180$. 

The question then becomes, given the constraints on ${\mathcal V}_p$ and ${\mathcal V}_6$, how small can $f_a$ be when compactifying on non-factorizable manifolds?  For example, in App.~\ref{sec:swiss_cheese} we discuss a class of non-factorizable compactifications with $p = 4$ in which $x \sim {\mathcal V}_4^2 / (g_s^2 {\mathcal V_6})$, such that the right hand side of~\eqref{eq:DwedgeD} is independent of the dimensionless volumes, in the limit ${\mathcal V}_6 \gg 1$. In this case, $f_a \sim M_s / g_s \sim M_{\rm pl} / \sqrt{ {\mathcal V}_6}$, such that $f_a$ cannot be brought parametrically below the string scale but can be numerically different than in~\eqref{eq:f_a_orbifolgGUT}.  In the KS examples discussed in the following section, on the other hand, we find cases where $f_a$ can be lowered even more while maintaining string scales above the GUT scale.

Along these lines, we now revisit the unitarity argument presented in Sec.~\ref{sec:unitarity} from the perspective of axions arising from $p$-form fields in string theory. 
In particular, we note that the smallest value of $M_{\rm KK}$, defined as the mass of the lowest SM KK mode, at fixed ${\mathcal V}_p$, is given by the case where the $p$-cycle is only stretched in one dimension and is of order the string scale in the other $p-1$ directions. This motivates the bound $M_{\rm KK} \gtrsim 2 \pi M_s / {\mathcal V}_p$.  Combining this relation with~\eqref{eq:unitarity_0} and using~\eqref{eq:key_relations} then leads to the lower bound 
\es{eq:f_a_volume_unitarity}{
f_a \gtrsim {\alpha_s^2(M_{\rm KK}) \over \sqrt{2} \pi} {M_{\rm pl} \over \sqrt{{\mathcal V}_6}} \,,
}
where the fine-structure constant is evaluated at the renormalization group (RG) scale $M_{\rm KK}$.  As we show in Sec.~\ref{sec:KS_axiverse}, all of our explicit KS constructions satisfy this bound, even those that do not allow for field theoretic unification. This bound also has the same parametric scaling with $M_{\rm pl}$ and ${\mathcal V}_6$ as that derived in~\cite{Reece:2024wrn} using perturbativity and naturalness arguments. More broadly, it is conceivable  that~\eqref{eq:f_a_volume_unitarity} is not the strongest possible bound from partial-wave unitarity.   
In particular, the theory should be unitary at the string scale, which may exceed the KK scale by a factor as large as $\mathcal{O}(1/\alpha_s)$. If the KK scale and string scale are separated, then a stronger unitarity bound may be derived applying the partial-wave unitarity arguments discussed in Sec.~\ref{sec:unitarity} at the string scale, accounting for the KK modes.\footnote{There are at least two complications to this calculation: (i) above the KK scale one must account for the additional multiplicity of gluon and axion KK modes, and (ii) above the KK scale the power-law running of the gauge couplings should be accounted for~\cite{Dienes:1998vh}. To partially account for (i), we may 
scatter in and out states in the KK-number singlet state   $\ket{\psi}_\mathrm{KK}=\sum_n\ket{G^{n}G^{n}}/\sqrt{N_\mathrm{KK}}$ (with $n$ the KK-number), 
mediated by an axion zero-mode. This  
would enhance the amplitude by the number of gluon KK modes with masses below $M_s$, $N_\mathrm{KK}\sim g_s/\alpha_\mathrm{UV}$, giving a unitarity bound on $f_a$ which has the same $\alpha_\mathrm{UV}$ dependence as \eqref{eq:matt_generic_Ms}.}  We leave such investigations to future work. (Note that in the warped case there may be subtleties associated to the theory being unitarized by a strongly coupled regime instead of by string states.)

Equation~\eqref{eq:f_a_volume_unitarity} shows that there is a lower-bound on the axion decay constant $f_a$, for a given total volume ${\mathcal V}_6$ and UV QCD gauge coupling $\alpha_s$, independent of unification. It is also phenomenologically interesting to ask if there is an upper bound on $f_a$. In our explicit KS constructions, without enforcing unification, we find no examples with $f_a \gtrsim 2 \times 10^{17}$ GeV.   
In the 5D orbifold construction this is intuitive.  Equation~\eqref{eq:key_relations} implies that $M_{s} \lesssim M_{\rm pl} / \sqrt{4 \pi}$, with the inequality saturated when all length scales are of order the string scale. In this case, all dimensionless cycle volumes are order unity; in the 5D case, for example, this implies $f_a = M_s / \sqrt{2 \pi}$ (see~\eqref{eq:decay_const_5D_flat_stringy_orbifold}). Taking $M_s \approx M_{\rm pl} / \sqrt{4 \pi}$ and $f_a \approx M_s / \sqrt{2 \pi}$ for the 5D case that gives rise to the largest $f_a$ then suggests the upper bound $f_a \lesssim M_{\rm pl} / (2 \sqrt{2} \pi) \sim 3 \times 10^{17}$ GeV for all axions, not just the QCD axion. A similar argument should hold in the full theory. On the other hand, and as we discuss in more depth later in this work, we expect perturbative and non-perturbative corrections which are not accounted for to be important in the small-volume limit~\cite{Conlon:2006tq}, so this upper bound $f_a \lesssim 3 \times 10^{17}$ GeV should be treated with caution.

We note that this upper limit on $f_a$ is weaker than that conjectured in the \textit{electric} 0-form Weak Gravity Conjecture (WGC). 
This conjecture (see, {\it e.g.},~\cite{Banks:2003sx,Arkani-Hamed:2006emk} and \cite{Reece:2024wrn} for a recent review) implies, for an $\mathcal{O}(1)$ constant $c$,\footnote{To our knowledge, there are no examples in the literature saturating this inequality with $c$ larger than $\sqrt{3/2}$. An example with $c=\sqrt{3/2}$ is given in Ref.~\cite{Reece:2024wrn}.} that $f_a \le c\mpl/S_\mathrm{inst} \approx 1.5c \times 10^{16}$ GeV, assuming $S_\mathrm{inst}=2\pi/\alpha_\mathrm{UV}$ and $\alpha_\mathrm{UV}=1/25$.  (Our conjectured upper bound is essentially equivalent to that of the WGC if one takes $\alpha_{\rm UV} = 1$ for the latter, or if the instanton corresponds to a different kind of stringy instanton such as worldsheet instantons.) The WGC bound is explicitly violated (for $c \sim  1$) 
by some compactifications in the KS axiverse that we discuss. However, there are subtleties in applying the electric WGC to these compactifications. Firstly, the above formulation of the axion WGC assumes a single axion and the KS compactifications all have more than one axion; secondly, it is unclear whether bare contributions to the instanton potential should be accounted for in the discussion (see~\cite{Sheridan:2024vtt} for a discussion).  More broadly, achieving $f_a \gtrsim 3 \times 10^{17}$ GeV may be possible, even for the QCD axion, in the case of highly anisotropic compact manifolds with multiple axions that undergo non-trivial mixing, though we are unable to engineer any examples with this behavior or find any explicit realizations in the KS axiverse (admittedly, though, our search might not be complete). 

\section{QCD axion from the Kreuzer-Skarke axiverse}
\label{sec:KS_axiverse}

Let us now turn to the KS axiverse construction in type IIB string theory, where axion decay constants and masses may be computed exactly across a wide variety of topologically non-trivial compactifications. We work in a weakly coupled limit of type IIB string theory, and compactify on O3/O7 orientifolds of CY 3-fold hypersurfaces of toric varieties. The CY 3-folds are obtained by triangulating reflexive polytopes of dimension 4 (using the construction of \cite{Batyrev:1993oya}), which have been enumerated in the KS database  \cite{Kreuzer:2000xy} and result in compactifications with up to $491$ axions. We study closed string axions arising from the dimensional reduction of $C_4$ over a basis of prime toric divisors of the integral homology group $H_4$.

The KS-axiverse differs from the toy models discussed previously in a number of important ways. For example, the typical decay constants $f_a$ in a given KS compactification decrease with increasing number of axions $N_a$, such that for $N_a\sim \mathcal{O}(1)$,  $f_a\sim \,10^{16}$ GeV  
while for $N_a \gtrsim 100$, $f_a \lesssim \, 10^{10}$ GeV.   
The large $N_a$ effect is due to a positive correlation between the number of homologically distinct 4-cycles in a CY 3-fold, which is equal to $N_a$, and the overall volume of the manifold. 

\begin{figure}[!htb]
    \centering
    \includegraphics[width=0.5\textwidth]{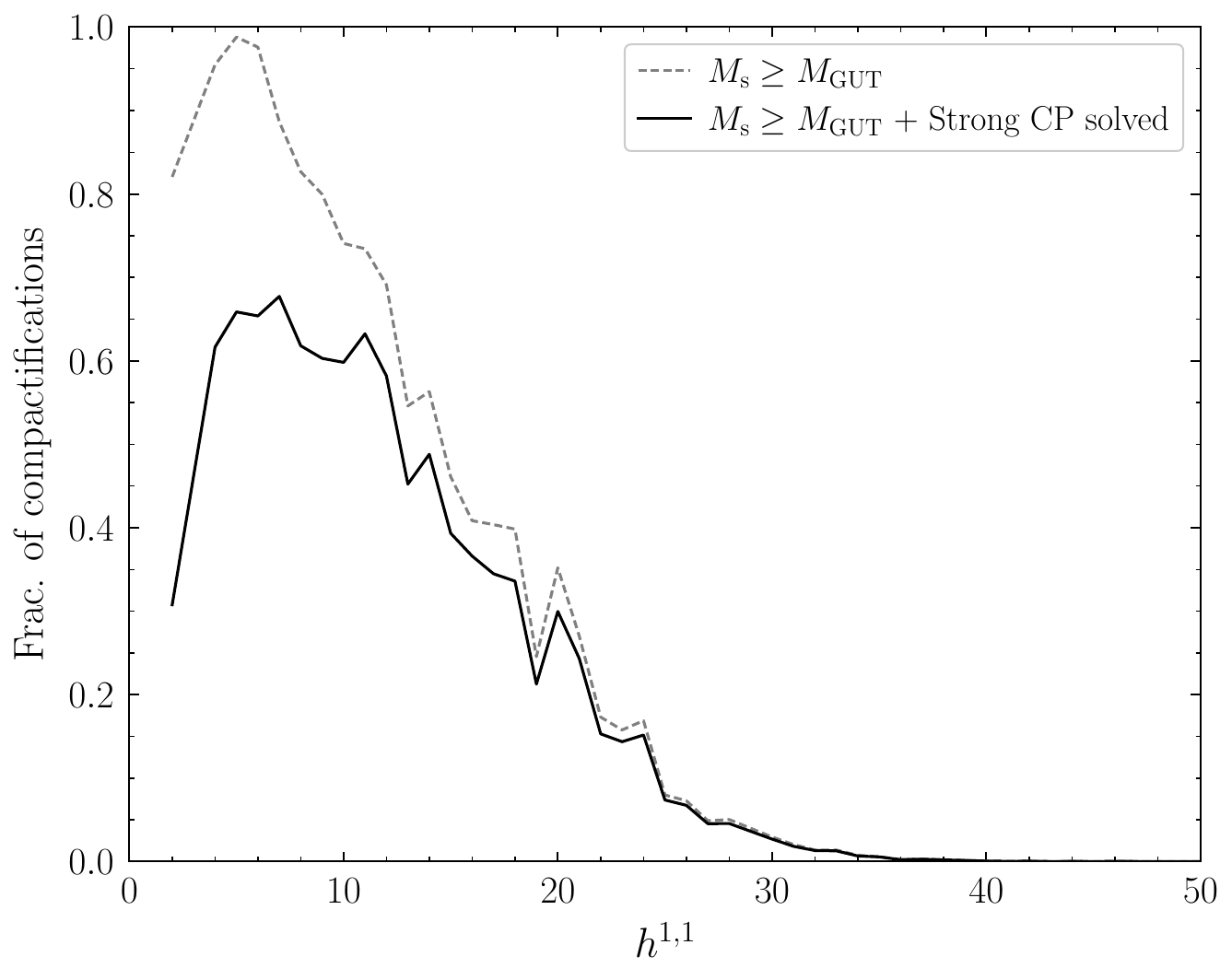}
    \caption{
    The fraction of KS compactifications for which $M_\mathrm{s} \gtrsim M_\mathrm{GUT}$ (gray), and the fraction for which this condition is satisfied and the Strong $CP$ problem is also solved ({\it i.e}, consistent with the neutron electron dipole moment constraint on the QCD vacuum angle, ($|\bar{\theta}|\lesssim 10^{-10}$) 
    (black), varying the number of axions $h^{1,1}$. 
    }
    \label{fig:mQCD_dist_field_theoretic_GUT}
\end{figure}

To construct ensembles of KS axiverses, we follow closely the procedure described in~\cite{Gendler:2023kjt}, which relies heavily on machinery developed in Refs. \cite{Kreuzer:2000xy, Demirtas:2018akl, Demirtas:2021gsq, Demirtas:2018akl, Gendler:2023hwg, Mehta:2021pwf}.  In particular, we make use of the package \texttt{CYTools} \cite{Demirtas:2022hqf} to sample CY 3-folds and study their topological data, which can be directly translated to data of the 4D axion EFT, as we summarize in App.~\ref{app:KS_details} (also see \cite{Demirtas:2018akl,Demirtas:2022hqf}). In brief, the decay constants and masses of the axion eigenstates depend on the Kähler metric, the overall volume of the manifold, and the volume of the basis divisors. We compute these data at a specific point in the Kähler moduli space, the tip of the \textit{stretched Kähler cone} \cite{Demirtas:2018akl}. We operate under the assumption that the EFT data does not depend strongly on the particular choice of location in the moduli space. 
This assumption was tested numerically to some extent in~\cite{Demirtas:2022hqf}, and in more detail in Ref.~\cite{Sheridan:2024vtt}, where possible caveats were discussed. 

For each $h^{1,1} \le 60$, we sample up to $100$ ``favorable" polytopes and $500$ \textit{fine, regular, star triangulations} per polytope, generating $\sim 1.7 \times 10^6$ compactifications in total.\footnote{We~use~the~\texttt{CYTools} function \texttt{random\_triangulations\_fast}. Note that while this algorithm does not necessarily generate a fair sampling of the triangulations, 
our calculation of the \textit{range} of QCD axion masses consistent with field-theoretic unification does not require one.} For each triangulation, we select a prime toric divisor at random to host QCD. To reproduce the observed value of the IR gauge coupling of QCD, we homothetically rescale the tip of the stretched Kähler cone to adjust the volume of the divisor $D_\mathrm{QCD}$ hosting QCD so that $\mathrm{\mathrm{Vol}}(D_\mathrm{QCD})=25$ in string-length units. (We set $g_s = 1$ and discuss the $g_s$ dependence of our results later in this section.)  
The manifold is discarded if any basis divisor has volume less than unity in string units. 
 
We show the fraction of compactifications for which $M_\mathrm{s} \ge M_\mathrm{GUT}$, as a function of $h^{1,1}$, in  Fig. \ref{fig:mQCD_dist_field_theoretic_GUT}.  We find that this condition is not satisfied for any of the sampled compactifications with $h^{1,1}\gtrsim 48$.  This has important implications for the KS axiverse, as it implies that the number of axion-like particles is at most 47 in these constructions. The distribution of QCD axion masses for the compactifications for which $M_\mathrm{s} \ge M_\mathrm{GUT}$ is shown in Fig.~\ref{fig:money}.\footnote{Note that while the histogram boundaries in Fig.~\ref{fig:money} are physically relevant, the shape of the histogram within the boundaries is strongly prior dependent. For this figure we use a prior where each $h^{1,1}$ value is equally likely, for definiteness.} We conclude that imposing $M_s > M_{\rm GUT}$ implies that the QCD axion mass satisfies $m_a \lesssim\, 10^{-8 }\,\mathrm{eV}$.

As a lower bound on the axion mass, we find that all of our compactifications have $f_a \lesssim 2 \times 10^{17}$ GeV, corresponding to $m_a \gtrsim 3 \times 10^{-11}$ eV, in agreement with the analytic arguments presented in Sec.~\ref{sec:p-form_axion}.  However, we caution that the high-$f_a$ examples have low total volume ${\mathcal V}_6$, as may be seen in Fig.~\ref{fig:fa_vol_scatter_plot}.  Corrections to the Kähler potential, for example in the $\alpha'$ expansion, are suppressed in the large ${\mathcal V}_6$ limit but could become important at small ${\mathcal V}_6$~\cite{Conlon:2006tq}. Since these corrections are not accounted for in this work, the low-${\mathcal V}_6$ points, corresponding to our cases with the largest $f_a$'s, are not necessarily reliable.  Thus, our lower QCD axion mass prediction $m_a \gtrsim 3 \times 10^{-11}$ eV should be treated with some degree of suspicion.

For the compactifications with $M_s \gtrsim M_{\rm GUT}$, we may also verify that they contain a sufficiently high-quality QCD axion to solve the Strong {\it CP} problem to adequate accuracy ($|\bar \theta| \lesssim 10^{-10}$ \cite{Abel:2020pzs,Peccei:2006as} with $\bar \theta$ the physical SM theta angle). Indeed, compactifications with sufficiently small divisor volumes generate stringy instantons that may shift the minimum of the QCD axion potential by more than one part in $10^{10}$. If these contributions do not shift the QCD vacuum angle $\bar{\theta}$ by more than $10^{-10}$, then the Strong $CP$ problem is solved. We verify that for the KS compactifications with $M_s > M_{\rm GUT}$, the QCD axion generically solves the Strong {\it CP} problem to sufficient accuracy if the only non-perturbative contribution to the axion potentials is from stringy instantons, as shown in Fig. \ref{fig:mQCD_dist_field_theoretic_GUT} (see App. \ref{app:KS_details} for details and Ref.~\cite{Demirtas:2021gsq}, where this question was first addressed in the KS axiverse).

\subsection{Axion string tension conjecture}
Let us comment on the connection between our results and the conjecture of Ref.~\cite{Reece:2024wrn} that, given an extra-dimensional axion with decay constant $f_a$, 
there exists an associated axion string with core tension
\begin{equation}
    \mu_* \approx 2\pi S_\mathrm{inst.} f_a^2\,,
\end{equation}
where $S_\mathrm{inst.}$ is the corresponding instanton action.\footnote{Here, by core tension we mean the tension of the string ignoring the contribution which is logarithmically divergent in the IR from the axion field configuration.} 
The conjecture was shown in  Ref. \cite{Reece:2024wrn} to hold in a number of string theory constructions. For example, in the case of an axion descending from $C_4$ in type IIB string theory, there is an associated axion string in the form of a D3 brane wrapped on the 2-cycle dual to the cycle on which $C_4$ is dimensionally reduced, of tension $\mu \sim \pi f_a \mpl$ (in the absence of warping)~\cite{Benabou:2023npn}. The instanton action is $S_\mathrm{inst.}\sim 2\pi/\alpha_\mathrm{GUT}\sim \mpl/f_a$, such that the tension takes the form specified previously. 

In Ref.~\cite{Reece:2024wrn} it is conjectured that in general, the axion string tension is expected to be at or above $\sim \Lambda_\mathrm{QG}^2$, with $\Lambda_\mathrm{QG}\sim M_s$ the quantum gravity cutoff of the theory. At least in the case of a string formed from a wrapped D$p$-brane, this can be shown explicitly. The tension of such a string is $\mu=T_p V_{\Omega}$, with $V_{\Omega}$ the volume of the ($p-1$)-cycle $\Omega$ which is wrapped, and $T_{\mathrm{D} p}=2 \pi M_s^{p+1} / g_s$ the tension of the D$p$-brane. In order for the $\alpha'$ expansion to be under control, $V_{\Omega}$ must not have a volume smaller than $\sim l_s^{p-1}$, from which it follows that $\mu \gtrsim T_\mathrm{F}/g_s$ with $T_\mathrm{F}=2\pi M_s^2$ the tension of the fundamental (F) string. In the weakly-coupled limit where $g_s <1$, we therefore have $\mu \gtrsim 2\pi M_s^2$.

The fact that the axion string tension $\mu_*$ lies above $\Lambda_\mathrm{QG}^2$ implies 
\begin{align}
\Lambda_\mathrm{QG} \lesssim  \sqrt{2\pi S_\mathrm{inst.}} f_a \,.
\label{eq:axion_string_tension_conjecture}
\end{align}
 In particular, for a QCD axion whose instanton action is from QCD gauge instantons, $S_\mathrm{inst.}=2\pi/\alpha_\mathrm{UV}$. While the precise value one should take for $\Lambda_\mathrm{QG}$ is not obvious, it is reasonable to take $\Lambda_\mathrm{QG}\approx M_s$, which is where the supergravity theory breaks down \cite{Green:1987sp}, and is also the first resonance of the open string in type IIB (note the first resonance of the closed string is at $2M_s$).  
In summary, from the string tension conjecture of Ref.~\cite{Reece:2024wrn} one obtains
\begin{align}
f_a \gtrsim \frac{\sqrt{\alpha_\mathrm{UV}}}{2\pi}M_s  \,,
\label{eq:matt_generic_Ms}
\end{align}
though we emphasize that the relation in Ref.~\cite{Reece:2024wrn} is only claimed to hold up to ${\mathcal O}(1)$ numbers.  
Note that this inequality would bound $f_a/M_s$ from below,
 similarly to the constraint \eqref{eq:f_a_volume_unitarity} following from the unitarity of the 4D axion EFT up to the scale $\sqrt{s}=M_\mathrm{KK}$. 

\begin{figure}
    \centering
    \includegraphics[width=1.0\linewidth]{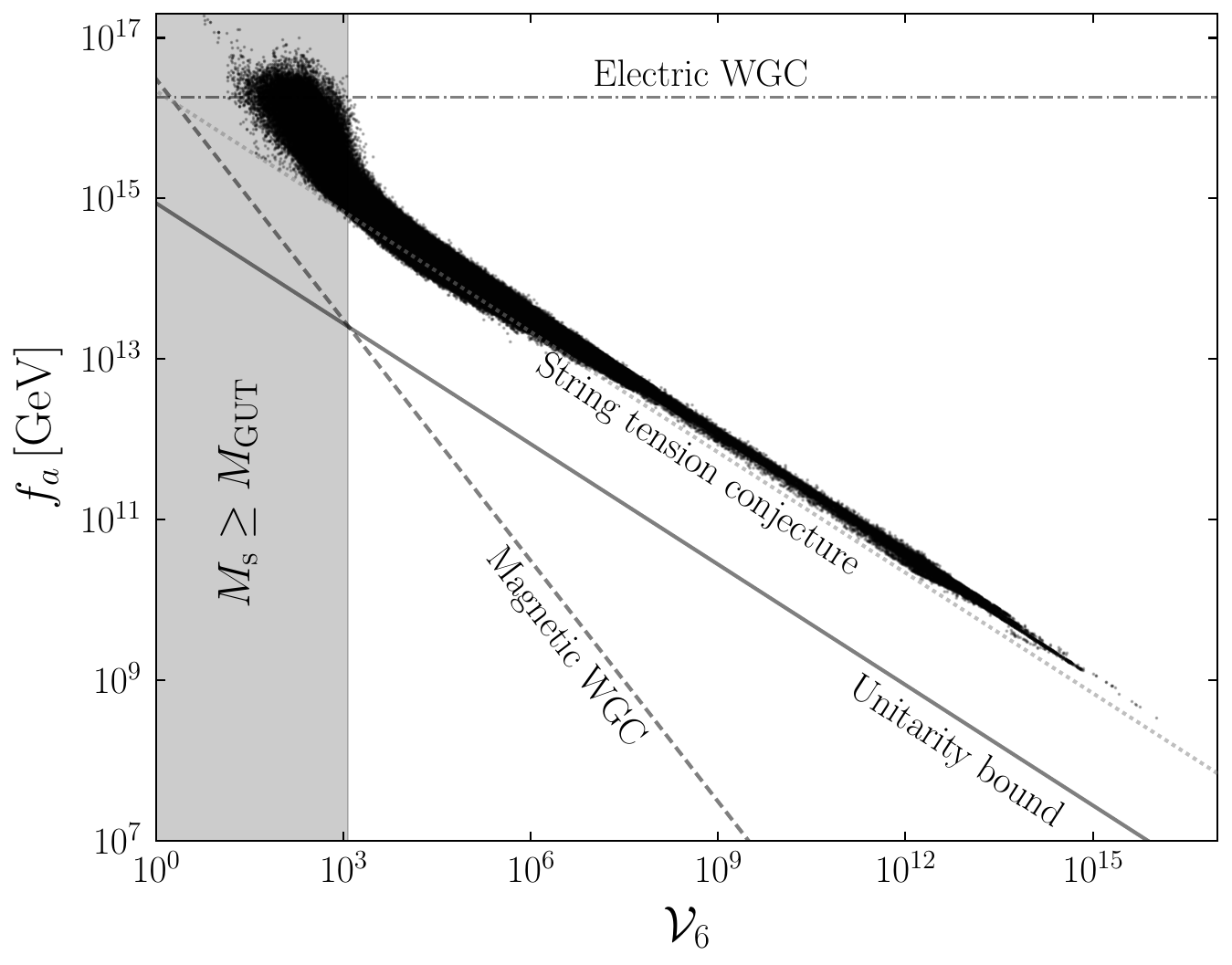}
    \caption{The joint distribution QCD axion decay constants $f_{a}$, which are defined through the axion-QCD interaction~\eqref{eq:axion_QCD}, and manifold volumes $\mathcal{V}_6$ (in units of $l_s$) for $\sim$$1.7 \times 10^6$ independent KS compactifications (points).  Note that the QCD axion decay constants here are independent of $g_s$. The KS compactifications obey, roughly, the scaling $f_a \propto \mathcal{V}_6^{-1/2}$. The range of volumes ${\mathcal{V}}_6$ consistent with grand unification is shaded in gray.   
    Compactifications obeying the conjectured axion string tension bound \eqref{eq:axion_string_tension_conjecture} (0-form \textit{magnetic} WGC) lie above the dotted (dashed)  line, though we stress the WGC bound applies for single axions where $f_a$ is the periodicity of the axion field and are thus not directly applicable to this parameter space, while the string tension conjecture is approximate. Unitarity of the 4D axion EFT \eqref{eq:f_a_volume_unitarity} imposes that $f_a$ lie above the  solid line (see Sec. \ref{sec:p-form_axion} for details); this bound does apply directly to our parameter space and is obeyed by all of the KS compactifications. Lastly, we indicate the maximal decay constant allowed by the \textit{electric} 0-form WGC (assuming $c=\sqrt{3/2}$, see text), $f_a =  1.8 \times 10^{16}$ GeV.  However, we caution that the WGC bound also need not apply to our multi-axion scenarios with $f_a$ defined through the axion-QCD interaction; indeed, it is violated by a number of our explicit compactifications.  
    For the purpose of illustration, all lines and shaded regions are plotted assuming $g_s=1$ (see text for details).   
    }
    \label{fig:fa_vol_scatter_plot}
\end{figure}

Combining the conjectured inequality~\eqref{eq:matt_generic_Ms} and the condition $M_\mathrm{s} \ge M_\mathrm{GUT}$ implies the QCD axion mass is bounded by 
\begin{align}
m_a \lesssim 9 \times 10^{-9} \, \mathrm{eV} \left(\frac{\alpha_\mathrm{UV}^{-1}}{25}\right)^\frac{1}{2} \,.
\label{eq:ma_bound_matt}
\end{align}
As we show in Fig.~\ref{fig:fa_vol_scatter_plot}, the inequality \eqref{eq:matt_generic_Ms} is satisfied  for the vast majority of our sampled KS compactifications, with a small fraction of compactifications violating it by a factor of at most $2$ in $m_a$. 
Note that the bound \eqref{eq:axion_string_tension_conjecture} is generally much stronger than the 0-form \textit{magnetic} WGC, $\Lambda_\mathrm{QG} \lesssim \sqrt{2\pi f_a \mpl}$, which follows from bounding the axion string tension above by $2\pi f_a\mpl$ (see \cite{Heidenreich:2021yda} for details), and which is easily satisfied by all of the KS compactifications we sample. It is also stronger than the unitarity constraint \eqref{eq:f_a_volume_unitarity}, which we also show in Fig.~\ref{fig:fa_vol_scatter_plot}. Note that this constraint assumes $M_\mathrm{KK}$ may be as low as $(2\pi\alpha_s)M_s$, while  
it is also possible that $M_\mathrm{KK}\sim M_s$, which would raise the lower bound on $f_a$ at fixed $\mathcal{V}_6$ by an $\mathcal{O}(1)$ factor. The unitarity constraint shown in Fig.~\ref{fig:fa_vol_scatter_plot} could therefore be made stronger in some UV completions.  

We caution that the WGC and the string tension conjecture need not apply in a straightforward way to our KS constructions for a few reasons.  
First, we emphasize that the string tension conjecture is only approximate~\cite{Reece:2024wrn}. Second,  
the string tension and WGC bounds define $f_a$ through the periodicity of the axion field, while for us $f_a$ is defined through the coupling of the axion to QCD as in~\eqref{eq:axion_QCD}. These two definitions are not equivalent in the case of multiple axions with non-trivial mixing.\footnote{A more detailed discussion of the difference between the two definitions of $f_a$ in the simplified two axion case can be found in \cite{Fraser:2019ojt}.}

Lastly, let us clarify the dependence of quantities plotted in Fig.~\ref{fig:fa_vol_scatter_plot} on $g_s$. First, for a fixed CY 3-fold in the KS database, the decay constant of the associated QCD axion  is independent of $g_s$ (see App.~\ref{app:KS_details} for details). To compute the decay constant we work in Einstein frame, {\it i.e.} with 4-cycles volumes $\tau$ defined in units of $g_sl_s^4$, such that the QCD divisor has $\tau=\alpha_\mathrm{GUT}^{-1}$. In these units the overall volume of the manifold is $\tilde{\mathcal{V}}_6\equiv\mathcal{V}_6/g_s^{3/2}$, with $\mathcal{V}_6$ the volume in units of string length. Compactifications with $M_s>M_\mathrm{GUT}$, for which proton decay constraints are respected and field theoretic unification is allowed, have volumes smaller than a critical volume: $\tilde{\mathcal{V}}_6\lesssim \tilde{\mathcal{V}}_{6*}$. The maximal volume $\tilde{\mathcal{V}}_{6*}$ scales with the string coupling as $g_s^{1/2}$. It follows that the minimal value of $f_a$ compatible with field-theoretic unification scales very mildly with the string coupling as $g_s^{-1/4}$ (since we observe that $f_a \propto 1/\sqrt{\tilde{\mathcal{V}}_6}$). To conservatively bound this value from below, we set $g_s=1$ in Fig.~\ref{fig:fa_vol_scatter_plot} (of course, in reality $g_s < 1$ is required for perturbativity). Similarly, the magnetic form of the WGC and the axion string tension conjecture of Ref.~\cite{Reece:2024wrn} give a lower bound on $f_a$ which scales as $g_s^{1/4}$. The electric form of the axion WGC is independent of $g_s$.

\section{Discussion}
\label{sec:discussion}
In this paper we derive mass range predictions for the QCD axion in scenarios where the axion arises as the zero mode of a closed-string-sector gauge field in string theory.  We impose $M_s\gtrsim M_{\rm GUT}$ to accommodate gauge unification and most naturally avoid proton decay. 
We find that the decay constant cannot be parametrically lowered from the string scale in flat-space compactifications, leading to an upper bound on the QCD axion mass (see~\eqref{eq:main}). In warped compactifications, on the other hand, the axion decay constant may be warped down to scales parametrically smaller than the string scale, though in these cases the KK scale is also warped to scales parametrically similar to $f_a$. These cases are thus inconsistent with the usual picture of SUSY grand unification at $M_{\rm GUT}$, and additional model building---such as localization---is required to avoid proton-decay-generating operators at the KK scale. In the case of heterotic M-theory on a warped interval with the SM on the smaller boundary, one can obtain a low $f_a$ at the price of lowering the M-theory scale $M_{11}$, which leads also to non-GUT phenomenology at high energies. We understand these claims through example constructions and, more generally, by partial-wave unitarity arguments.

A number of loopholes remain that could allow for QCD axion masses above the neV scale. Below, we enumerate two of these possibilities:

1) One of the simplest realizations where the relevant scales can be separated include theories with anomalous $U(1)$'s where the axion comes from the mixing between a higher-form axion and a phase of a complex scalar, $\phi$. In these theories there can be cancellations in the D-term potential for $\phi$, $V_D\propto \left ( |\phi|^2+\sum_i q_i\frac{\partial K}{\partial \tau_i}\right )^2$, leading to field theoretic axions with low decay constant, $f_a\sim |\phi| \ll M_s$, as recently explored in ~\cite{Choi:2011xt,Cicoli:2013ana,Cicoli:2013cha,Buchbinder:2014qca,Choi:2014uaa,Allahverdi:2014ppa, Loladze:2025uvf,Petrossian-Byrne:2025mto}. This kind of cancellation may disentangle the axion decay constant from $M_s$ allowing for heavier axions with $m_a\gtrsim 10^{-7}$ eV, which can be probed in laboratory experiments such as ADMX, HAYSTAC, MADMAX, ALPHA, or BREAD~\cite{Adams:2022pbo}. Interestingly, this scenario is also a promising way to obtain axion strings for stringy axions \cite{Loladze:2025uvf,Petrossian-Byrne:2025mto}. The cancellations involved at the level of the D-term potential resemble the required cancellations in 4D GUTs when the vacuum expectation value (VEV) of the Higgs breaking the PQ symmetry is much smaller than the Higgs VEV breaking the GUT symmetry.

2)  
Of course, another possibility---within the context of closed-string axions---is simply to accept a string scale or KK scale below $M_{\rm GUT}$. In this case, one could imagine a compact space which is very large $\mathcal{V}_6\gg 1$ leading to a low string scale $M_s\ll \mpl$ or, alternatively, keeping $M_s > M_{\rm GUT}$ but warping down $f_a$ and $M_{\rm KK}$.  
Minimal realizations of this idea are challenging to construct, however,  and will be typically at odds with proton decay bounds~\cite{Super-Kamiokande:2020wjk}, unless additional model building is incorporated ({\it e.g.},~\cite{Svrcek:2006yi,Heckman:2024trz}). Specifically, one is challenged to explain why the KK modes of GUT gauge bosons, if grand unification occurs at a scale near $M_{\rm KK}$, do not lead to fast proton decay as discussed in~\cite{Hebecker:2001wq}. One should forbid operators induced by stringy physics of the type 
\begin{equation}
    \mathcal{L} \supset c\frac{1}{M_s^2}qqq\ell\,,
\end{equation}
for quark $q$ and lepton $\ell$ fields.
Building models that avoid these two issues is possible but requires ingredients that make the theory non-minimal and \textit{less unified}; {\it e.g.}, by separating SM fermions that would belong to the same GUT multiplet and locating them at different points in the compact space~\cite{Arkani-Hamed:1999ylh,Aldazabal:2000cn,Gherghetta:2000qt} in order to decrease the proton decay rate to acceptable values. As shown in~\cite{Cvetic:2009ng,Cvetic:2010mm}, these contributions are calculable and can play a relevant role in constraining some bottom-up approaches in intersecting D-brane constructions. 

Lastly, let us briefly comment on our conclusions in the context of the recent DESI DR2 results, which suggest an evolving dark energy (DE) equation state with $w$ growing in time~\cite{DESI:2025zgx,Lodha:2025qbg}. (Note that the DESI DR1 data already showed a preference for evolving DE~\cite{DESI:2024aqx,DESI:2024kob}.) Ref.~\cite{Lodha:2025qbg} finds that DESI DR2 DE data are consistent with an axion interpretation, where an ultralight axion with mass $m_a\sim H_0$ undergoes misalignment beginning close to the hilltop of its potential $V(\varphi)=m_a^2 f_a^2\left[1+\cos \left(\varphi / f_a\right)\right]$, with initially $0.7 \lesssim \varphi/f_a \lesssim 1$. In this case, we may have $w \sim -1$ at early times, with $w$ growing  as the axion rolls towards the minimum of its potential. In this scenario, a joint likelihood over CMB, DESI, and three supernovae datasets (PantheonPlus, Union3, and DESY5) constrains 
the decay constant to be $\log_{10} \left(f_a / \mpl\right)=-0.13_{-0.29}^{+0.33}$ for PantheonPlus, $-0.29_{-0.35}^{+0.63}$ for Union3, and $-0.09_{-0.40}^{+0.66}$ for DESY5~\cite{Lodha:2025qbg}. 

The axion interpretation of the DESI data is, however, in tension with an extra-dimensional axion UV completion for a few reasons. Foremost, it is difficult to obtain the required size for the axion decay constant while also keeping the axion mass small, $m_a \sim 10^{-33}$ eV.\footnote{Note that here, unlike for the QCD axion, we define the axion-like particle decay constant through the periodicity of the canonically normalized axion field---in particular defining the axion to have periodicity $2 \pi f_a$---in the limit of weak kinetic mixing. In contrast, for the QCD axion we define $f_a$ through the axion-QCD interaction. In practice, to compute the upper limit on $f_a$ for low-mass axions in our KS construction we select $m_a \lesssim H_0$ and compute the axion periodicities  ignoring kinetic mixing. } In all of the extra-dimensional axion constructions discussed in this work, we have $f_a \lesssim 2 \times 10^{16}$ GeV for axions with masses $m_a \lesssim H_0$, while $f_a \gtrsim 2.5 \times 10^{17}$ GeV is needed to explain the evidence for evolving DE. This result is also consistent with the electric WGC~\cite{Arkani-Hamed:2006emk} (see discussion below Eq.~\eqref{eq:f_a_volume_unitarity}) and is supported by  
the examples studied in Apps. \ref{sec:5D_orbifold_flat}, \ref{sec:p-form_axion}, and heterotic axions. 
 Physically, since typical axion potentials generated by stringy instantons are given by $V(a)\propto  M_s^4e^{-S}\cos (a/f_a)$, the required size for $f_a$ to explain the DE data is in strong conflict with the simultaneous need for the axions to be light ($m_a \sim H_0$)~\cite{Banks:2003sx}. (In the presence of light fermions, such as in low-scale SUSY, this qualitative expectation can be modified.)  On the other hand, large effective axion decay constants have been constructed previously in string theory contexts~\cite{Banks:2003sx,Abe:2014pwa,Shiu:2015xda,Bachlechner:2014gfa}, and we cannot rule out the possibility that through careful engineering a combination of multiple axion fields ({\it e.g.},~\cite{Kaloper:2005aj,Svrcek:2006hf,Reig:2021ipa}) could produce the required behavior to explain the DESI data.

\section*{\textbf{Acknowledgments}}
We thank Prateek Agrawal, Itay Bloch, Michele Cicoli, Nathaniel Craig,  Lawrence Hall, James Halverson, Arthur Hebecker, Anson Hook, Igor Klebanov, Soubhik Kumar, David Marsh, John March-Russell,  Jakob Moritz, Matthew Reece, Nick Rodd, Eva Silverstein, and Edward Witten for useful discussions. We are especially grateful to Naomi Gendler and Liam McAllister for many useful discussions. 
J.B. and B.R.S. are supported in part by the DOE award
DESC0025293. B.R.S. acknowledges support from the
Alfred P. Sloan Foundation. KF is supported by the Miller Institute for Basic Research in Science, University of California Berkeley, and also thanks the Aspen Center for Physics, which is supported by NSF grant PHY-2210452, for hospitality while working on this project.
This research used resources of the National Energy Research
Scientific Computing Center (NERSC), a U.S. Department of Energy Office of Science User Facility located
at Lawrence Berkeley National Laboratory, operated under Contract No. DE-AC02-05CH11231 using NERSC
award HEP-ERCAP0023978.

\bibliography{Bibliography}

\clearpage
\appendix

%%%%%%%%%%%%%%%%%%%%%%%%%%%%%%%%%%
    
\section{Warped extra dimension}
\label{sec:warped}
One of the original motivations for warped extra dimensions is explaining the smallness of the weak scale~\cite{Randall:1999ee}. In the presence of a warped interval, with branes at the endpoints,  the IR brane cut-off is warped down from the fundamental Planck scale $M_5$ (which in explicit string constructions will be related to $M_s$) as $\Lambda_\mathrm{IR}\sim M_5 e^{-k\pi R}$, while the KK scale is defined as $M_{\rm KK}\sim k e^{-k\pi R}$. In typical Randall-Sundrum (RS) scenarios, where warping is employed to solve the hierarchy problem, one has $M_5 \sim \mpl$ and  $kR\sim \mathcal{O}(10)$ so that $\Lambda_{\rm IR}$, which defines the cut-off at the IR brane, is close to the TeV scale. 
In this section we will see that, for a QCD axion coming from a bulk gauge field, both the KK scale and the decay constant will receive the same warp factor $\sim e^{-k\pi R}$. In minimal scenarios this will make $f_a$  an intermediate scale between the IR cut-off and the KK scale, $M_\mathrm{KK}\lesssim f_a \lesssim \Lambda_\mathrm{IR}$.

In warped extra dimensions, the 4D Planck scale is defined in terms of the size of the interval, the AdS curvature and the string scale as\footnote{Note that in the RS literature, the 4D Planck scale is typically related to the 5D Planck scale as $\mpl^2=\frac{M_5^3}{k}\left(1-e^{-2 \pi k R}\right)$. To facilitate the comparison with the stringy literature we make the string coupling and string scale explicit in these formulas.}
\begin{equation}
    \mpl^2=\frac{2\pi}{g_s^2}\frac{M_s^3}{k}\left(1-e^{-2 \pi k R}\right)\,.
\end{equation}
Let us now study axions from the fifth component of an abelian gauge field $A_M$ in a theory with a warped extra dimension~\cite{Choi:2003wr,Flacke:2006ad}. The 5D action, including the CS coupling of the gauge field to the GUT gauge group (with field strength $G_{MN}$) is 
\begin{align}
S_{5D} &\supset \int d^4x \, dy \, \sqrt{-G} \left( -\frac{1}{2 g_5^2} A_{MN} A^{MN} \right. \nonumber \\
&\quad \left. + \frac{\kappa}{16\pi^2} \frac{1}{\sqrt{-G}} A_M G_{NP}^a \tilde{G}^{MNP\,a} \right) \,.
\end{align}
Note that the quantization of $\kappa$ is independent of the background metric.  Assuming $A_5$ does not vary in the 5th dimension, it satisfies 
\begin{align}
A_{\mu 5}=\partial_\mu A_5-\partial_5 A_\mu=\frac{ \pi k R e^{2 k y}}{e^{2 \pi k R}-1} \partial_\mu A_5 \,,
\end{align}
such that upon dimensional reduction we obtain
\begin{align}
S_{\mathrm{axion}}^{4D} = \int d^4x \Bigg[ -\frac{1}{2}\frac{\pi^2 k R^2}{2 g_5^2} \frac{1}{e^{2 \pi k R}-1} \partial_\mu A_5 \partial^\mu A_5 \nonumber \\
\quad + \frac{\kappa_F}{16\pi^2} \pi R A_5 G_{\mu\nu}^a \tilde{G}^{a\,\mu\nu}\,\, \Bigg] \,.
\end{align}

From this we obtain the axion decay constant
\begin{equation}\label{eq:decay_constant_RS}
    f_a^2 = \frac{k}{g_5^2\left ( e^{2k\pi R} -1\right )}=\frac{1}{g_s} \frac{k M_s}{\left ( e^{2k\pi R} -1\right )} \,,
\end{equation}
where in the last equality we have used that, as in the flat space case, $\frac{1}{g_5^2}=\frac{M_s}{g_s}$. For $\pi\alpha_{\rm GUT} \lesssim g_s \lesssim 1$, and AdS curvature comparable to the string scale, $k\approx M_s\gg R^{-1}$, 
the decay constant goes as 
\begin{equation}\label{eq:decay_const_warped_GUT}
    f_a = e^{-k\pi R}\sqrt{\frac{kM_s}{g_s}}= \frac{M_{\rm KK}}{\sqrt{\alpha_{\rm GUT}}}\frac{1}{\sqrt{k\pi R}}\,.
\end{equation}
where in the last equality we anticipated using the definition of the gauge coupling in the warped extra dimension,~\eqref{eq:gauge_coupling_RS}. 

We note that unlike the flat extra dimension case, the decay constant is exponentially sensitive to the extra dimension size. This warp factor may be an issue in cases where the bulk is GUT symmetric, e.g. a warped 5D orbifold GUT. The reason is that the gauge coupling at the compactification scale is given by
\begin{equation}\label{eq:gauge_coupling_RS}
    \frac{1}{\alpha_{\rm GUT}} = \frac{\pi R}{g_5^2}=\frac{1}{g_s}\frac{\pi R}{l_s}\,.
\end{equation}
For $k\approx M_s$, we see from ~\eqref{eq:decay_constant_RS} and~\eqref{eq:gauge_coupling_RS} that the decay constant is not separated from the KK scale (which in this case lies close to the IR brane cut-off), which for $ \alpha_{\rm GUT}^{-1}\sim kR = \mathcal{O}(10)$ (as required to solve the hierarchy problem) would imply $f_a \sim M_{\rm KK} \sim \mathcal{O}(\text{TeV})$. Such a low decay constant is excluded by  e.g, neutron star cooling, which bounds  $f_a \gtrsim 3 \times 10^8$ GeV (with only a minor dependence on the details of the UV completion) \cite{Buschmann:2021juv}. 

A way to raise $f_a$ to be within non-excluded parameter space 
is to separate the AdS curvature from the string scale\footnote{This and other related issues may also be solved in more sophisticated multi-throat constructions~\cite{Flacke:2006ad}.}, $k\ll  M_s$, which in turn implies $M_{\rm KK}\ll \Lambda_{\rm IR}$. 
For the sake of completeness let us consider an example with $kR=3$, $R^{-1}\approx M_{\rm GUT}$, and $\alpha_{\rm GUT}=1/25$. In this case, one obtains $f_a=8\times 10^{12}$ GeV, which leads to a QCD axion mass around $m_a=7\times 10^{-7}$ eV. We see that in the presence of warping, we can enhance the axion mass with respect to the prediction in the flat case. 

However, as discussed in the main text, this introduces two issues. First, the unification of gauge couplings will be typically spoiled due to running in the 5D theory. Second, one has to deal with proton decay. If the SM fermions were located at or near the IR brane, the proton decay rate would be too large since $M_{\rm KK}\ll M_{\rm GUT}$. One can employ localization mechanisms to suppress the dimension-6 operator $\sim \frac{c}{\Lambda^2}qqql$ that leads to proton decay. This operator can be induced by either stringy physics or by KK modes of GUT gauge bosons. Regarding the former, as shown in~\cite{Gherghetta:2000qt}, the RS scenario allows for exponential suppression of the coefficient $c$ if the SM fermions are placed towards the Planck brane. In that case, $\frac{c}{\Lambda^2}\sim \frac{1}{\mpl^2}$ and the dimension-6 operator may be sufficiently suppressed. In~\cite{Pomarol:1999ad,Pomarol:2000hp} it was shown that the KK mode mediated proton decay is also suppressed as we locate the SM fermions towards the UV brane. This occurs because the wavefunction of the GUT gauge bosons peaks towards the IR brane.

We conclude that if the visible (GUT) gauge sector propagates in a warped extra dimension with the axion coupled to gauge bosons through a CS term, the AdS length has to be larger than the fundamental string length and close to size of the extra dimension, $kR\sim \mathcal{O}(1)$. In this case, the effect of warping allows to decouple the decay constant $f_a$ from $R^{-1}$ parametrically and $3\times 10^8 \text{ GeV}\lesssim f_a \lesssim f_a^{\rm flat}$ is possible. Since the value of $kR$ is not fixed by 4D low-energy observables such as gauge couplings, the warped axion case is less predictive than its flat-space counterpart (see \eqref{eq:decay_const_5D_flat_stringy_orbifold}).

\section{Simple non-factorizable examples}
\label{sec:swiss_cheese}

We now consider a type IIB-like scenario where the GUT symmetry comes from wrapping D7 branes over 4-cycles and axions come from $C_4$ integrated over 4-cycles. Let us take a toy 
 ``Swiss-cheese" model with 3 moduli fields in the limit where one of the cycles is much larger than the rest, $\tau_1\gg\tau_{2,3}\sim \alpha_{\rm GUT}^{-1}$, so that the visible sector (GUT) can be placed on either of the two smaller cycles \cite{Conlon:2006tq,Reece:2024wrn}.\footnote{Note that here, in a similar way to Sec.\ref{sec:KS_axiverse}, we work in the Einstein frame where factors of $g_s$ are absorbed into the definition the 4-cycle volumes $\tau$.} 

The total volume is given in terms of the real parts of the complexified 4-cycle moduli, $T_j=\tau_j+ia_j$, as
\begin{equation}
\label{eq:swiss_V6}
    \mathcal{V}_6= (T_1+\bar{T}_1)^{3/2}-(T_2+\bar{T}_2)^{3/2}-(T_3+\bar{T}_3)^{3/2}\,.
\end{equation}
The Kähler potential is given in terms of the volume as  $\mathcal{K}=-2\ln\mathcal{V}_6$
and the decay constants can be obtained from the Kähler metric, $\mathcal{K}_{ij}= 2\frac{\partial^2 \mathcal{K} }{\partial T^i \partial \bar{T}^j}$. 
For large volumes $\mathcal{V}_6\gg 1$, the Kähler metric is diagonal to good approximation (see \cite{Conlon:2006tq} for the exact form)
\begin{align}
  \mathcal{K}_{ij} \sim \begin{pmatrix}
\frac{-3}{2\sqrt{2\tau_1}\mathcal{V}_6}+\frac{9\tau_1}{\mathcal{V}_6^2} & \mathcal{V}_6^{-5/3} & \mathcal{V}_6^{-5/3} \\
\mathcal{V}_6^{-5/3} & \frac{3}{2\sqrt{2\tau_2}\mathcal{V}_6}+\frac{9\tau_2}{\mathcal{V}_6^2} & \mathcal{V}_6^{-2} \\
\mathcal{V}_6^{-5/3} & \mathcal{V}_6^{-2} & \frac{3}{2\sqrt{2\tau_3}\mathcal{V}_6}+\frac{9\tau_3}{\mathcal{V}_6^2} \\
\end{pmatrix}\,,
\end{align}
and the decay constants are approximately given by the diagonal elements as
\begin{equation}
    f_{a_i}^2 \approx \frac{1}{16\pi^2} \mpl^2\mathcal{K}_{ii}\,.
\end{equation}

In the case of interest we place the GUT symmetry on the 2nd or 3rd cycle and the associated decay constant is

\begin{equation}
    f_{a_i} \approx \frac{1}{4\pi}\mpl\sqrt{\mathcal{K}_{ii}} \propto  \mpl / \sqrt{\mathcal{V}_6}\,,\,\,\text{for } i=2,3\,.
    \label{eq:swiss_cheese_fa_vol}
\end{equation}

Therefore $f_a$ is expected to have a scaling similar to the string scale up to order $O(1)$ factors. Similar scalings have been obtained more recently in~\cite{Cicoli:2022fzy}. 
Note that this scaling differs from the expectation for an approximately isotropic manifold studied above in Sec.~\ref{sec:p-form_axion}, for which $f_a$ and the overall volume of the manifold are fixed. 
More generally, we find that scanning over the full possible range of $\tau_1$, $\tau_2$, and $\tau_3$ in this geometry the QCD axion decay constant can be as small as $f_a \approx 2 \times 10^{15}$ GeV, corresponding to $m_a \approx 3 \times 10^{-9}$ eV, while keeping $M_s \gtrsim M_{\rm GUT}$.

Conversely, we note that low QCD axion masses may be obtained here by carefully adjusting the $\tau_i$ values so that ${\mathcal V}_6$ is small because of a cancellation between the positive and negative terms in~\eqref{eq:swiss_V6}. For example, consider the case where $\tau_1 = 40$, $\tau_2  = 25$, and $\tau_3 = 25$, so that ${\mathcal V}_6 \approx 8$. The QCD axion decay constant in this scenario is around $f_a\approx 10^{18}$ GeV.  While this large decay constant violates the bound we conjecture in Sec.~\ref{sec:p-form_axion}, we emphasize that the corrections to the Kähler metric are likely not under control in this example. For example, Ref.~\cite{Conlon:2006tq} argues that the Kähler potential corrections should be suppressed by $\sqrt{\tau_{2,3}} / {\mathcal V}_6$ in the Swiss-cheese model, which means that they are unsuppressed since this ratio is of order unity. Thus, we cannot claim to have a reliable example with $f_a > 10^{18}$ GeV for the QCD axion, though we can also not rule out the possibility that such an example exists once the full Kähler potential is accounted for.

Another beautiful example of theories where the decay constant scales independently of the gauge coupling comes from type IIB where the compact manifold $X_6$ is obtained from the conifold after replacing the singularity by $S^2$~\cite{Svrcek:2006yi}. In this case the gauge sector is obtained from D5-branes wrapping the vanishing cycle $S^2$ and the gauge coupling depends on the size of the $S^2$ as $\alpha_{\rm GUT}=g_s \mathcal{V}_{S^2}^{-1}$, with $\mathcal{V}_{S^2}$ the volume of $S^2$ in string units. The axion comes from the RR 2-form, expanded as $C_2=\frac{a}{2\pi}\frac{\omega_2}{4\pi}$, with $\omega_2$ a harmonic 2-form on $X_6$. After expressing $\omega_2$ and its Hodge dual $\star\omega_2$ in terms of a basis of 1-forms on $X_6$, it can be shown that $\int_{X_6}\omega_2\wedge\star\omega_2\propto \mathcal{V}_6^{1/3}$ up to $\mathcal{O}(1)$ factors which depend on details about the manifold and how the singularity is resolved.
In this case, the decay constant is independent of the gauge cycle volume and given by
\begin{equation}
    f_a^2 \approx \frac{g_s^2}{24\pi}\frac{\mpl^2}{\mathcal{V}_6^{2/3}}\,. 
\end{equation}

\section{Computations in the Kreuzer-Skarke axiverse}
\label{app:KS_details}
In Type-IIB orientifold compactifications on the manifold $X_6$, the decay constants of axions descending from the RR-form $C_4$ are determined by the homology group $H_4(X, \mathbb{Z})$, which may be decomposed into a basis of $h^{1,1}$ divisors $\left\{\left[D_a\right]\right\}$. The Kähler form on $X_6$ is parametrized in terms of curve volumes $t_a$ as
\begin{align}
J=\sum_{a=1}^{h^{1,1}} t_a\left[D_a\right] \,.
\end{align}
In this section we work in Einstein frame, where we write the volume of divisor $\Sigma_4$ as $\tau=\mathrm{Vol}(\Sigma_4)/(g_sl_s^4)$ with $\mathrm{Vol}(\Sigma_4)$ the dimensionful volume of $\Sigma_4$ (note that in the main text we do not absorb $g_s$ into $\tau$). The volume of $X_6$ and the four-cycle volumes $\tau^a$ can then be computed in terms of the Kähler parameters $t_a$ and the triple intersection form $\kappa^{a b c}\equiv\int_{X_6}\left[D_a\right] \wedge\left[D_b\right] \wedge\left[D_c\right]$, as
\begin{align}
\mathcal{V}_6&\equiv\int_{X_6} \frac{1}{3!} J \wedge J \wedge J=\frac{1}{3!} \kappa^{a b c} t_a t_b t_c, \\
\tau^a&\equiv\frac{1}{2} \int_{D_a} J \wedge J =\frac{1}{2} \kappa^{a b c} t_b t_c \,.
\end{align}
In the low-energy 4D EFT, there are $h^{1,1}$ closed-string axions obtained by dimensionally reducing $C_4$ over the $D_a$,
\begin{align}
\theta^a\equiv 2\pi\int_{D_a} C_4 .
\end{align}
Note that, combined with the four-cycle volumes $\tau^a$, which may be treated as real scalar fields, the axion is the imaginary component of a complex scalar in a chiral multiplet of $\mathcal{N}=1$ SUSY.

Dimensional reduction of the ten-dimensional gravitational action yields an axion kinetic term of the form:
\begin{align}
\mathcal{L}_{ \mathrm{kin. }}=-(2\pi)^{-2}\frac{M_{\mathrm{pl}}^2}{2} \mathcal{K}_{a b} \partial_\mu \theta^a \partial^\mu \theta^b \,,
\end{align}
where $\mathcal{K}_{a b}$ is the Kähler metric on Kähler moduli space. The Kähler metric $\mathcal{K}^{a b}$ is obtained by taking derivatives of the Kähler potential with respect to the 4-cycle volumes as:
\begin{equation}
    \mathcal{K}_{ab}=\frac{1}{2}\frac{\partial}{\partial\tau^a}\frac{\partial}{\partial\tau^b}\mathcal{K}\,,\,\,\text{ with: }\mathcal{K}=-2\log\mathcal{V}_6\,.
\end{equation}
In this work we use \texttt{CYTools} to compute $\mathcal{V}_6$, the $\tau_i$, and $\mathcal{K}_{a b}$ for CY 3-folds generated from the KS database.

\begin{figure*}[!t]
    \centering
    \includegraphics[width=0.48\textwidth]{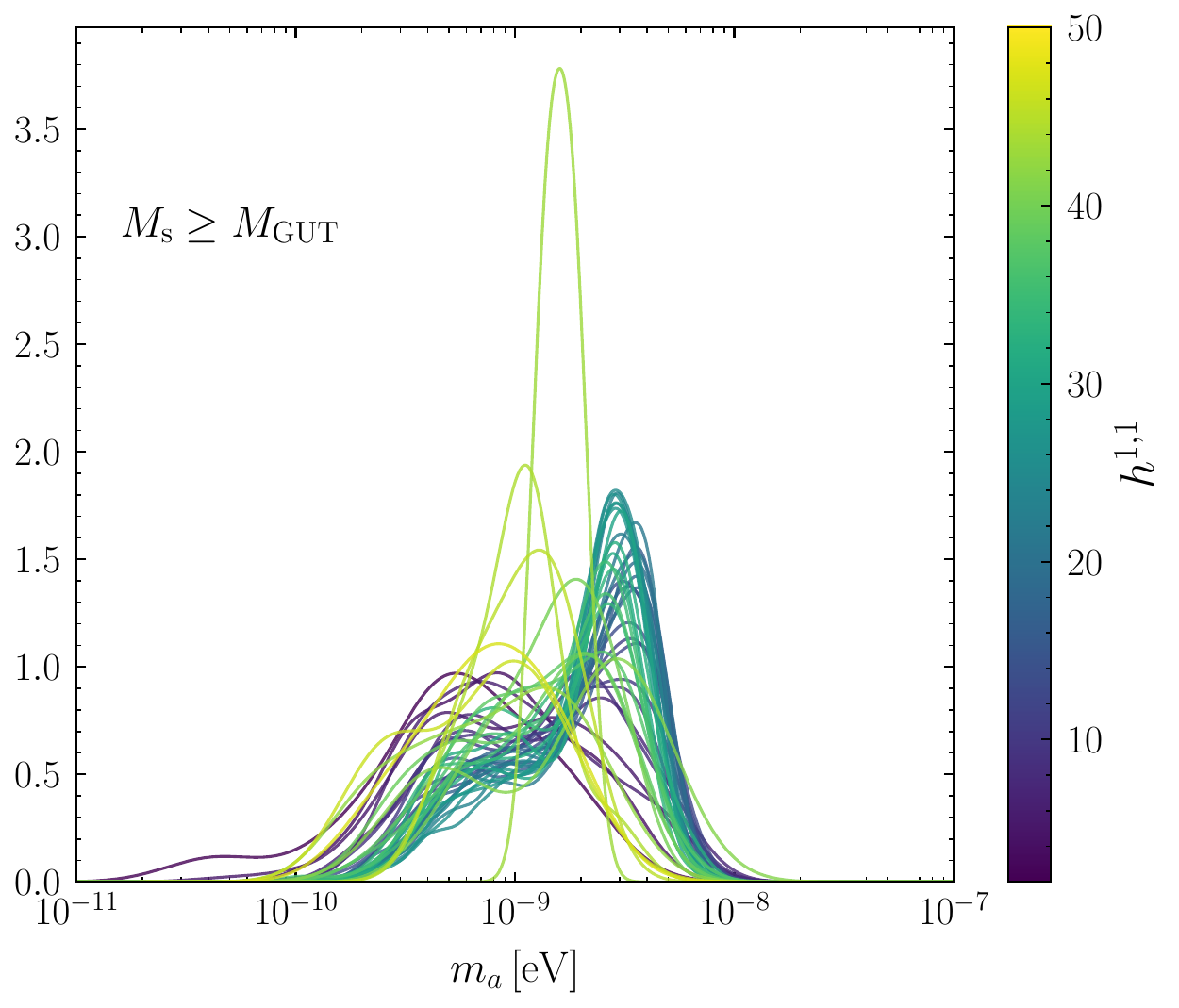}     \includegraphics[width=0.48\textwidth]{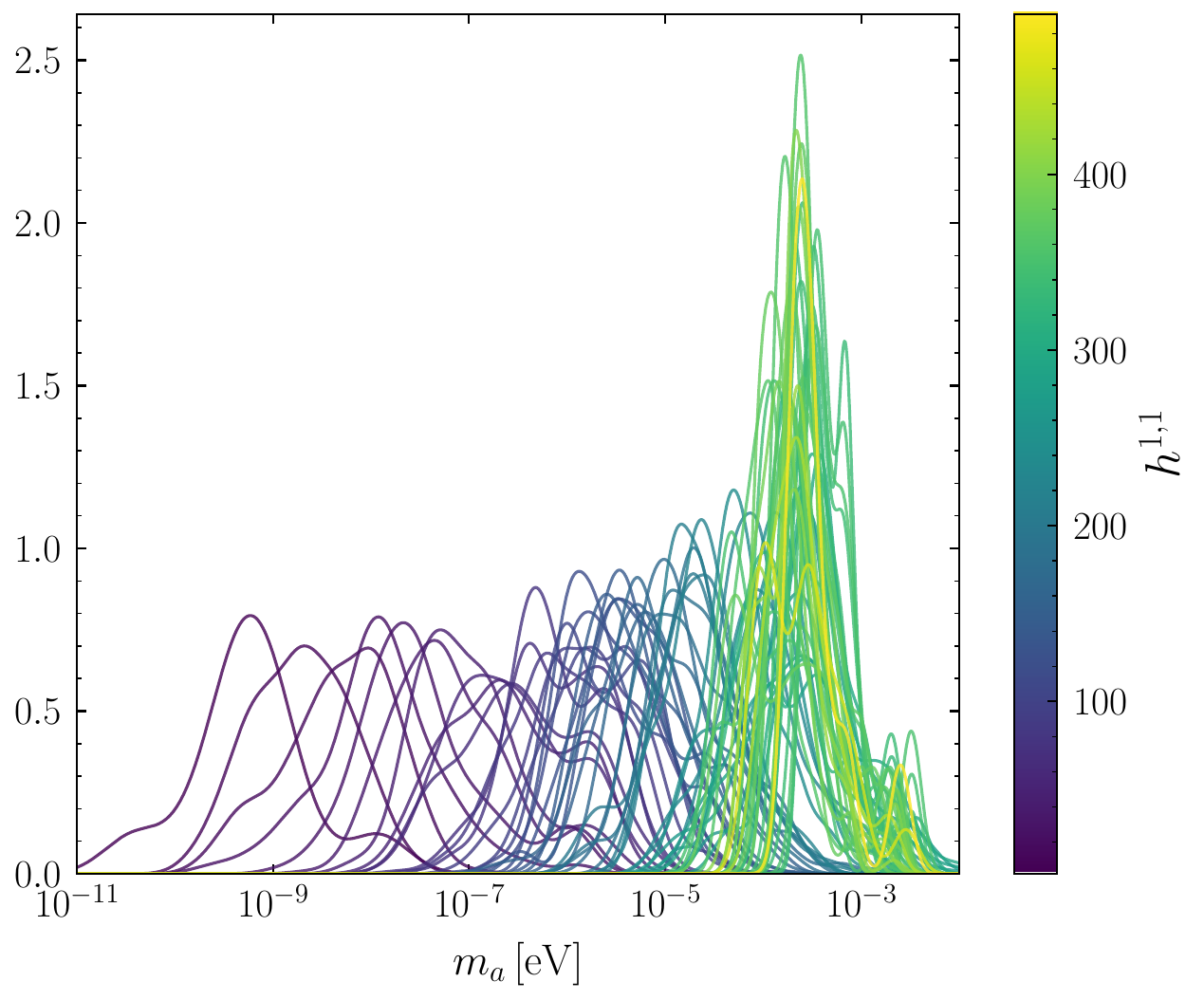}
    \caption{(\textit{Left}) For individual values of  $h^{1,1}$,  the distribution of the QCD axion mass at fixed $h^{1,1}$ (in the form of a kernel density estimate (KDE)), restricting to compactifications for which $M_\mathrm{s} \ge M_\mathrm{GUT}$ (see Fig.~\ref{fig:money} for the full distribution over all $h^{1,1}$). Note that in this plot the support of the KDE extends slightly beyond the smallest and largest $m_a$ realized in our samples.
    (\textit{Right}) KDEs of the distribution of QCD axion mass in the KS axiverse, varying the Hodge number $h^{1,1}$.
    }
    \label{fig:combined_ma_QCD_KS}
\end{figure*}

\begin{figure}
    \centering
    \includegraphics[width=1.0\linewidth]{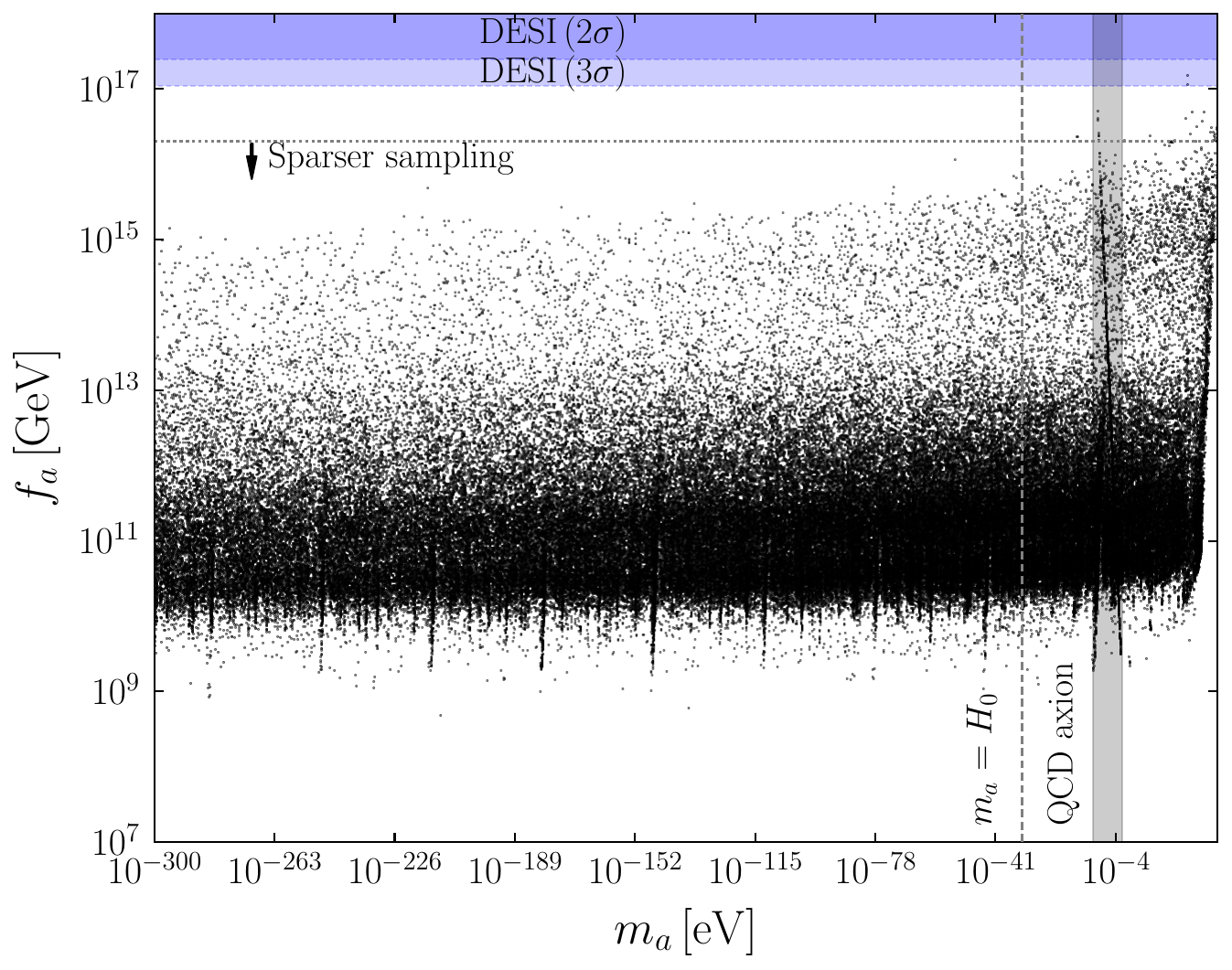}
    \caption{Including all sampled compactifications with the UV value of the QCD gauge coupling equal to $\alpha_\mathrm{GUT}=1/25$,  
    the joint distribution $(m_a, f_a)$ over all axions. For clarity, we plot more sparsely the sampled compactifications with $f_a \le 2 \times 10^{16}$ GeV. 
    Here we define the axion decay constant through the periodicity of the canonically normalized field, computed by ignoring off-diagonal kinetic mixing. We caution that this differs from the definition used in this work of the QCD axion decay constant, which is through the axion-gluon coupling. The 2 and 3 sigma bands consistent with an axion interpretation of the DESI DR2 data are shaded. 
    There are no sampled compactifications which are not in tension with this axion explanation.}
    \label{fig:joint_ma_fa_KS_axiverse}
\end{figure}

\begin{figure}
    \centering
    \includegraphics[width=1.0\linewidth]{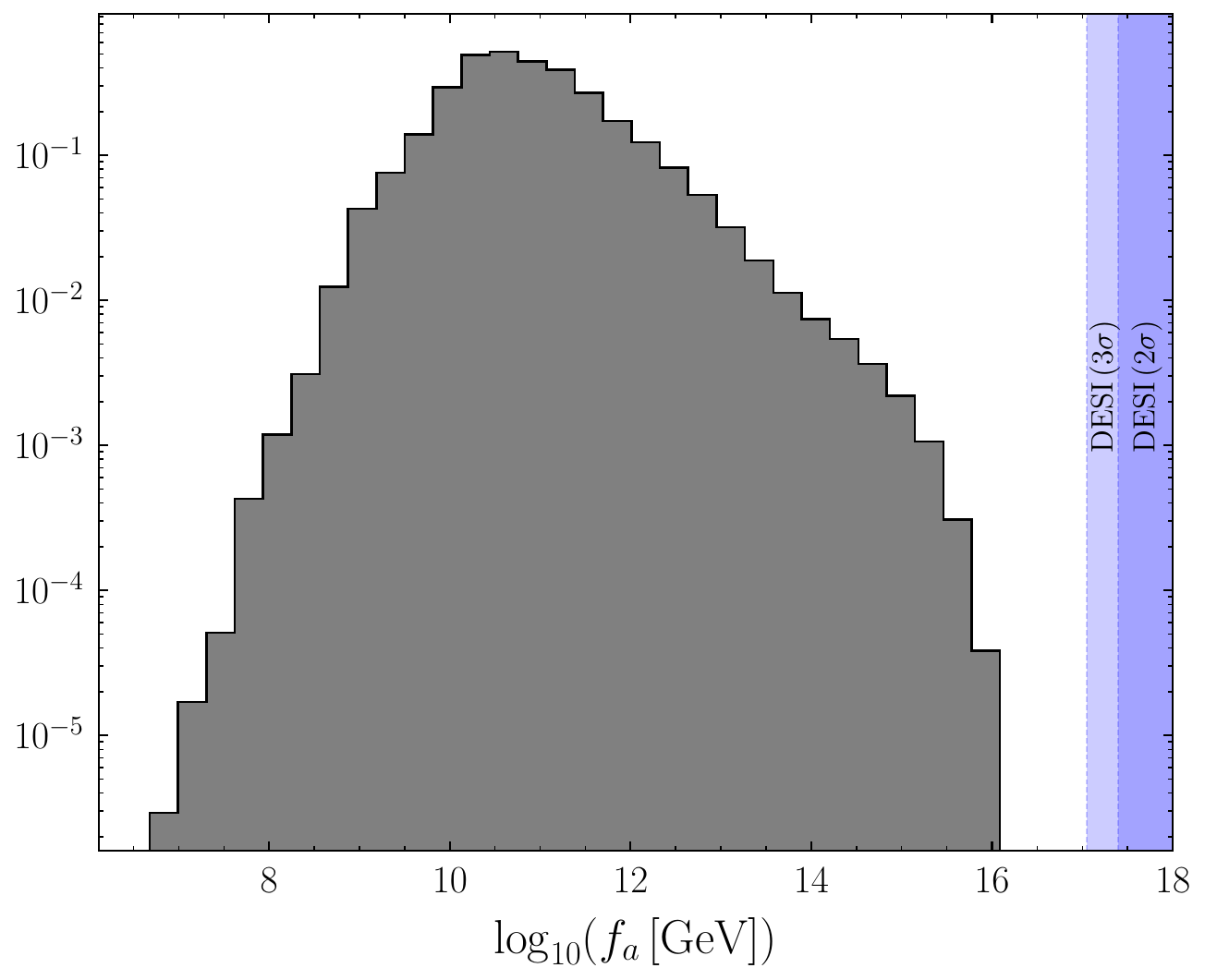}
    \caption{As in Fig.~\ref{fig:joint_ma_fa_KS_axiverse}, but for
    the distribution of $f_a$ for axion-like particles with $m_a \le H_0$.  
    As in Fig.~\ref{fig:money}, for each $h^{1,1}$ we reweight the distribution by the number of sampled compactifications of Hodge number $h^{1,1}$.  
    }
    \label{fig:fa_KS_axiverse_DESI}
\end{figure}

%%%%%%%%%%%%%%%%%%%%%%%%%%%%%%%%%%%%%%%%%%%%%%%%%%%%%%%%%%%%%%%%%
The shift symmetry of the axions is protected by the higher-form global symmetry of $C_4$, but can be broken by non-local effects.
In particular, the axion potential receives contributions from the superpotential\footnote{There also could be contributions from the Kähler potential, though we do not account for them as they are less well understood (but see \cite{Demirtas:2021gsq}).}, through Euclidean D3 branes wrapping effective divisors. The resulting potential is given by \cite{Gendler:2023hwg}
\begin{widetext}
\begin{equation}
\begin{aligned}
V(\boldsymbol{\theta}) & =-\mpl^4\frac{8 \pi}{\mathcal{V}_6^2}\left[\sum_\alpha\left(\boldsymbol{q}^{(\alpha)} \cdot \boldsymbol{\tau}\right) e^{-2 \pi \boldsymbol{q}^{(\alpha)} \cdot \boldsymbol{\tau}} \cos \left( \boldsymbol{q}^{(\alpha)} \cdot \boldsymbol{\theta}+\delta_\alpha\right)\right. \\
& \left.+\frac{1}{2} \sum_{\alpha \neq \alpha^{\prime}}\left(\pi \boldsymbol{q}^{(\alpha) \top} \boldsymbol{K}^{-1} \boldsymbol{q}^{\left(\alpha^{\prime}\right)}+\left(\boldsymbol{q}^{(\alpha)}+\boldsymbol{q}^{\left(\alpha^{\prime}\right)}\right) \cdot \boldsymbol{\tau}\right) e^{-2 \pi\left(\boldsymbol{q}^{(\alpha)}+\boldsymbol{q}^{\left(\alpha^{\prime}\right)}\right) \cdot \boldsymbol{\tau}} \cos \left(\left(\boldsymbol{q}^{(\alpha)}-\boldsymbol{q}^{\left(\alpha^{\prime}\right)}\right) \cdot \boldsymbol{\theta}+\delta_{\alpha, \alpha^{\prime}}\right)\right] \,,
\end{aligned}
\end{equation}
\end{widetext}
where  $\mathbf{q}^\alpha$ are instanton charges of effective divisors, $\delta_{\alpha}$, $\delta_{\alpha,\alpha'}$ are phases, and $\mathbf{K}$ is the kinetic mixing matrix. Note that the effective cone is generated by prime toric divisors, of which there are $h^{1,1}+4$ for a \textit{favorable} polytope. The instanton charges of the prime toric divisors  correspond to the entries of the GLSM charge matrix \cite{Demirtas:2022hqf}, which is computed by \texttt{CYTools}. We retain the $h^{1,1}$ leading instanton contributions to the potential (see \cite{Gendler:2023hwg} for details), which are by construction CP-preserving, i.e such that the phases may be rotated away by the $h^{1,1}$ axions.
In summary,  \texttt{CYTools} provides the topological data necessary to compute the kinetic and mass mixing matrices for the $h^{1,1}$ axions. In practice, computation of the decay constants and masses of the canonically normalized axion mass eigenstates by direct diagonalization can be numerically unstable for large $h^{1,1}$. Instead, these quantities may be computed, to excellent approximation in the limit of large instanton scale hierarchies, using the algorithm described in Ref.~\cite{Gendler:2023kjt}. Distributions of the QCD axion mass for a large sampling of KS compactifications are shown varying $h^{1,1}$ in Fig.~\ref{fig:combined_ma_QCD_KS}\footnote{Here we sample 50 values between $h^{1,1}=5$ and $h^{1,1}=332$, and include every $h^{1,1}>333$ up to the largest value in the KS database, $h^{1,1}=491$.  Note that for any  $h^{1,1}< 333$ there exists a CY 3-fold with Hodge number $h^{1,1}$, but this is no longer true for $h^{1,1}\gtrsim 333$, with large gaps in $h^{1,1}$ for which no CY exists.}.

To determine whether a given KS compactification solves the PQ-quality problem, we require that a shift $\Delta \theta $ in the minimum of the QCD axion potential induced by a CP-violating stringy instanton respects neutron EDM constraints, i.e $|\Delta \theta| \le 10^{-10}$. Such a shift can be approximated by 
\begin{align}
&|\Delta \theta| \approx \frac{\Lambda_{\cancel{\mathrm{CP}}}^4}{\Lambda_{\mathrm{QCD}}^4} \,,
\end{align}
with $\Lambda_{\mathrm{QCD}}=0.075 \, \mathrm{GeV}$ \cite{diCortona:2015ldu}, and $\Lambda_{\cancel{\mathrm{CP}}}$ the maximal CP-violating instanton scale (note that here by instanton scale we refer to a scale $\Lambda$ for which $\Lambda^4$ is the prefactor of some cosine appearing in $V(\boldsymbol{\theta}$)). 
Except for $h^{1,1}\lesssim 5$, the majority of the compactifications satisfying $M_s \ge M_\mathrm{GUT}$ solve the Strong $CP$ problem, as shown in Fig.~\ref{fig:mQCD_dist_field_theoretic_GUT}.

Lastly, as discussed in  Sec.~\ref{sec:discussion}, to assess the possibility of a stringy axion interpretation of the DESI DR2 evidence for evolving DE, we scan the KS axiverse for  ultralight axion-like particles with  $m_a\sim H_0$ and with decay constants $f_a \gtrsim 2.5 \times 10^{17}$ GeV. In this interpretation, a single axion mass eigenstate undergoing misalignment  is responsible for the evolution of the DE equation of state $w$ from $-1$ at early times to larger values as the axion relaxes towards its minimum (we do not consider multi-axion dynamics for which lower $f_a$ might be allowed). The relevant definition of the decay constant here is thus the periodicity of the canonically normalized axion field, which in practice we compute by setting off-diagonal terms of the kinetic mixing matrix $\mathcal{K}$ to zero.  
Assuming this definition, the joint distribution of $(f_a, m_a)$ varying over all $h^{1,1}$ is shown in Fig.~\ref{fig:joint_ma_fa_KS_axiverse}. Restricting to axions with $m_a \le H_0$, the distribution of $f_a$ is shown in  Fig.~\ref{fig:fa_KS_axiverse_DESI} (note that while the axion interpretation above requires $m_a\sim H_0$, in Fig.~\ref{fig:fa_KS_axiverse_DESI} we only impose $m_a \lesssim H_0$ to have sufficient statistics).

\section{Non-trivial $N_\mathrm{DW}$ from higher-level embedding}
\label{app:Ndw}
Here we examine how domain wall (DW) number $N_\mathrm{DW}>1$ may be obtained for QCD axions arising from the reduction of $p$-form fields. For concreteness, let us work in Type~IIB with spacetime being a product manifold $X_6 \times M_4$, with $M_4$ Minkowski space. Following the conventions in~\cite{Ibanez:2012zz}, we note that the 10d CS term coupling the RR-form $C_p$ (with even $p$) to the gauge field living on a D$(p+3)$ brane wrapping the $p-$cycle $\Sigma_p$ is
\begin{align}
 S_\mathrm{CS}\supset\mu_{p+3}\frac{(2\pi \alpha^\prime )^2}{2}\int_{\Sigma_p \times M_4} C_p \wedge \mathrm{Tr}(F \wedge F),
\end{align}
with $F$ the gauge field strength, and $\mu_{p+3}=M_s^{p+4}/(g_s(2\pi)^{p+3})$. Note that the CS level here is $\kappa=1$, which derives from the fact that the D$(p+3)-$branes have charge 1. In standard situations, as we consider throughout the main text, this leads to $N_\mathrm{DW}=1$. A domain wall number $N_\mathrm{DW}>1$ can be relevant, for example, when estimating the lower bound on $f_a$ from partial-wave unitarity in Sec.~\ref{sec:unitarity}.

One can easily obtain $N_\mathrm{DW}>1$ without changing the D-brane charges. The simplest way to achieve this is by embedding non-trivially the IR gauge group, let's say QCD, into the gauge groups which are obtained from the D-branes. 
Schematically, we consider a product group which is broken spontaneously to the diagonal subgroup:
\begin{align}\label{eq:symm_breaking_diagonal}
    G_1\times ...\times G_{N_\mathrm{DW}}\rightarrow G_{\rm diag} \equiv G_{\rm QCD} \,.
\end{align}
One way to realize this is by Higgsing with bi-fundamental Higgs fields in the 4D EFT, although higher-dimensional mechanisms are also conceivable. In this case, the generators of $G_{\rm diag}$ are given in terms of the original generators, $T^a_{\rm diag} = \sum^k_i T_i^a$. 

This has two major implications.
The first is the mentioned non-trivial DW number, which can be seen after dimensional reduction
\es{}{
  &\mu_{p+3} \frac{(2\pi\alpha')^2}{2}\int_{\Sigma_p \times M_4} C_p \wedge \mathrm{Tr}(F_i \wedge F_i) \\
  &\to  
  N_\mathrm{DW}{\alpha_s \over 8 \pi} {a \over F_a}  G^a_{\mu \nu} \tilde G^{a \, \mu \nu}\,,
}
with Einstein summation over $i$ and $F_i$ the field strength associated to $G_i$. 
In the equation above $F_a$ is the fundamental period of the axion, determined geometrically, so that $f_a=\frac{F_a}{N_\mathrm{DW}}$. We obtain a non-trivial DW number from the higher-level embedding of QCD. See~\cite{Agrawal:2022lsp} for higher-level of embedding in the context of GUTs.

The second implication is the presence of UV (fractional) instantons which may be relevant for IR axion physics~\cite{Agrawal:2017ksf,Gaillard:2018xgk,Fuentes-Martin:2019bue,Csaki:2019vte}. The gauge coupling matching condition at the SSB breaking scale, $\alpha^{-1}_{\rm diag}=\sum_i \alpha^{-1}_i$, tells us that, individually, each of the $G_i$ gauge groups is more strongly coupled than the QCD (IR) gauge group at the symmetry breaking scale. Let us assume, for simplicity, that all $G_i$ have a comparable gauge coupling. In this case, $\alpha_i (M_\mathrm{SSB}) = N_\mathrm{DW}\alpha_{\rm diag} (M_\mathrm{SSB})$, which implies that the UV instanton action is reduced by the non-trivial DW number,
\begin{equation}
    S_{\rm UV}=\frac{2\pi}{\alpha_i}=\frac{2\pi}{N_\mathrm{DW}\alpha_{\rm diag}}\,,
\end{equation}
with the gauge couplings evaluated at the symmetry breaking scale, see \eqref{eq:symm_breaking_diagonal}. If $\alpha_{\rm diag}\sim \alpha_{\rm GUT}$, then the situation just described might lead to a quality problem if the vacua of the different $G_i$ groups is not aligned. We leave a more detailed study on other possible mechanisms to generate a non-trivial DW number for future work.

\onecolumngrid

\end{document}